\documentclass[aip,jap,reprint]{revtex4-1}

\usepackage{graphicx}
\usepackage{dcolumn}
\usepackage{bm}
\usepackage{amsmath}

\begin{document}

\title{Water nanoelectrolysis: A simple model}

\affiliation{CINaM-CNRS Aix-Marseille Univ., Campus de Luminy, case 913, 13288 Marseille Cedex 9, France}

\author{Juan Olives}
\email{olives@cinam.univ-mrs.fr}

\author{Zoubida Hammadi}

\author{Roger Morin}

\author{Laurent Lapena}

\date{\today}

\begin{abstract}
A simple model of water nanoelectrolysis---defined as the nanolocalization at a single point of any electrolysis phenomenon---is presented. It is based on the electron tunneling assisted by the electric field through the thin film of water molecules ($\sim$0.3 nm thick) at the surface of a tip-shaped nanoelectrode (micrometric to nanometric curvature radius at the apex). By applying, e.g., an electric potential $V_1$ during a finite time $t_1$, and then the potential $-V_1$ during the same time $t_1$, we show that there are three distinct regions in the plane $(t_1,V_1)$: one for the nanolocalization (at the apex of the nanoelectrode) of the electrolysis oxidation reaction, the second one for the nanolocalization of the reduction reaction, and the third one for the nanolocalization of the production of bubbles. These parameters $t_1$ and $V_1$ completely control the time at which the electrolysis reaction (of oxidation or reduction) begins, the duration of this reaction, the electrolysis current intensity (i.e., the tunneling current), the number of produced ${\rm O_2}$ or ${\rm H_2}$ molecules, and the radius of the nanolocalized bubbles. The model is in good agreement with our experiments.
\end{abstract}

\maketitle

\section{Introduction}

Water electrolysis is used for hydrogen production\cite{Turner:2004, Zeng-Zhang:2010, An-etal:2017} and more generally for the production of bubbles and the study of their formation, dissolution, stability, acoustic properties, etc.,\cite{Fernandez-etal:2014, Luo-White:2013, Czarnecki-etal:2015} owing to the numerous applications, e.g., in medicine.\cite{Bloch-etal:2004, Qamar-etal:2010, Acconcia-etal:2013} Microbubbles produced by electrolysis are also used to manipulate a microobject.\cite{Li-Hu:2013} Very small electrodes, with diameter $<$ 1 $\mu$m, were used for the study of microbubble/substrate forces or the formation of nanobubbles.\cite{Donose-etal:2012, Luo-White:2013, German-etal:2016} An effective control of the micro/nanobubbles, e.g., concerning their localization and size, is crucial for all these studies and applications. Nevertheless, if the electrode is not of micro/nanometric size, microbubbles will generally appear everywhere on the electrode surface.  With our new method, called nanoelectrolysis, the production of microbubbles can be controlled and nanolocalized at a single point, namely, the apex of a tip-shaped electrode (with micrometric to nanometric curvature radius at the apex).\cite{Hammadi-etal:2013} By means of nanoelectrolysis, a strong control of the microbubbles is obtained: a single bubble can be immobilized (or moved to any point) in the liquid, at some distance from the apex of the electrode.\cite{Hammadi-etal:2016} 

In addition to the known electrocatalytic effect of a high surface area on the electrode,\cite{Stojic-etal:2003, Nikolic-etal:2010, Wang-etal:2014} our approach shows the importance of the nanostructure/nanogeometry of the electrode surface: by applying a low electric potential during a finite time (e.g., using an alternating potential), the production of bubbles can be nanolocalized at a single reaction site with a nanometric/micrometric curvature radius on the electrode.\cite{Hammadi-etal:2013} In the same way, nanometric heterogeneities (optically invisible) are potential reaction sites for bubble production (see Fig.~1 and Note~16 in Ref.~\onlinecite{Hammadi-etal:2013}), and the activity of macroscopic electrodes is probably due to the presence of many such uncontrolled heterogeneities. According to our approach, nanostructured electrode surfaces---with arrays of nanotips or nanopillars, i.e., sites with nanometric curvature radii---could probably improve the electrode activity. 

However, the main interest of nanoelectrolysis is to produce calibrated microbubbles at a single site, in a controlled way. Calibrated microbubbles are needed in various medical applications, e.g., as ultrasound contrast agents for capillary imaging, drug delivery, or blood clot lysis.\cite{Bloch-etal:2004, Lindner:2004, Hernot-Klibanov:2008, Acconcia-etal:2013} They are used to study ultrasounds--microbubbles interactions with applications to the detection and sizing of bubbles, e.g., for the prevention of decompression sickness (scuba diving and extra-vehicular astronaut activity) or the monitoring of liquid sodium coolant in nuclear reactors.\cite{Wu-Tsao:2003, Czarnecki-etal:2015, Buckey-etal:2005, Kim-etal:2000} The advantage of nanoelectrolysis over the microfluidics technique of production of bubbles\cite{Whitesides:2006} is that no surfactant (biologically harmful) is used and that arbitrary bubble production frequency (even a single bubble production\cite{Hammadi-etal:2016}) can be obtained. Calibrated microbubbles were recently produced from nanoelectrolysis combined with ultrasounds using tap water (non-chemically controlled) solution.\cite{Achaoui-etal:2017} In this paper, we show that calibrated microbubbles of any size can be obtained at a single site by nanoelectrolysis with a chemically controlled solution, by applying a suitable electric potential during a finite time.

Although electrolysis is classically described using the electric potential (which is constant on the whole surface of each electrode), nanoelectrolysis reveals the fundamental role of the electric field, which is higher at the apex of the electrode (where the curvature radius is very small). Nanoelectrolysis is caused by the electron tunneling through the thin film of water layers at the electrode/solution interface, assisted by this high electric field.\cite{Hammadi-etal:2013} In this paper, we present a general and simple model based on this tunneling and field effect, which is in agreement with the experiments and explains the various aspects of nanoelectrolysis, i.e., the nanolocalization of each electrolysis reaction (oxidation and reduction) and of the production of bubbles. It leads to a complete control of these reactions and the bubble production, at a single point, by means of the applied electric potential. The model applies to any type of electrolysis in aqueous solutions involving gas production (provided the electrode surface is not altered by solid deposition).

\section{The model}

\begin{figure}[htbp]
\begin{center}
\includegraphics[width=7cm]{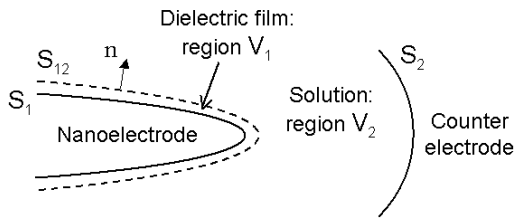}
\end{center}
\caption{The nanoelectrode, the dielectric film, the solution, and the counter electrode: general notations.} \label{Notations}
\end{figure}

\begin{figure}[htbp]
\begin{center}
\includegraphics[width=3cm]{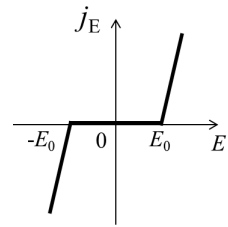}
\end{center}
\caption{Modeled dependence of the tunneling or electrolysis current $j_{\rm E}$ on the electric field $E$ in the dielectric film.} \label{E current}
\end{figure}

In our experiments, one of the two electrodes---called the nanoelectrode---is tip-shaped (and made of Pt), with a curvature radius, at the apex of the electrode, ranging from 5 $\mu$m to 1 nm.\cite{Hammadi-etal:2013, Hammadi-etal:2016} An aqueous solution of $\rm H_2SO_4$ ($10^{-4}$ to $10^{-3}$ mol/L) is generally used. The presence of a few water layers at the surface of the nanoelectrode\cite{Xia-Berkowitz:1995, Guidelli-Schmickler:2000, Rossmeisl-etal:2008, Osawa-etal:2008} will be modeled as a dielectric film of constant thickness $d$ ($d \approx$ 0.3 nm). Let us denote $\rm V_1$ as the region occupied by the dielectric film, $\rm V_2$ as that occupied by the solution, $\rm S_1$ the surface of the nanoelectrode, $\rm S_2$ that of the counter electrode, $\rm S_{12}$ the interface between $\rm V_1$ and $\rm V_2$, and $\rm n$ the unit vector normal to $\rm S_{12}$, oriented from $\rm V_1$ to $\rm V_2$ (Fig.~\ref{Notations}).
Maxwell's equations give the discontinuities of the normal components of the electric field $E$ and the electric current $j$ at the interface $\rm S_{12}$ 
\begin{align}
\varepsilon (E_{2\rm n}&-E_{1\rm n}) = \rho_{\rm s}\nonumber\\
j_{2\rm n}&-j_{1\rm n} = -\frac{\partial\rho_{\rm s}}{\partial t}\label{Maxwell}
\end{align}
(assuming the same permittivity $\varepsilon$ in $\rm V_1$ and $\rm V_2$; $\rho_{\rm s}$ is the surface charge density on $\rm S_{12}$), hence
\begin{eqnarray}
\gamma\,E_{2\rm n} = -\varepsilon\frac{\partial}{\partial t}(E_{2\rm n}-E_{1\rm n}) + j_{1\rm n},
\label{Interface}
\end{eqnarray}
with the help of Ohm's law $j_2 = \gamma\,E_2$ ($\gamma$ being the conductivity of the solution). The term $-\varepsilon\frac{\partial}{\partial t}(E_{2\rm n}-E_{1\rm n})$ represents a ``charging current''---more precisely, the current due to the discharge of the interface $\rm S_{12}$, according to Eq.~(\ref{Maxwell})---and will be denoted $j_{\rm C}$. Owing to the low thickness of the dielectric film, electrons can cross this film by quantum tunneling if the electric field in the film is high enough, producing the electric current $j_{1\rm n}$ responsible for the electrolysis reactions. This tunneling current $j_{1\rm n}$ will thus be called the electrolysis current and denoted $j_{\rm E}$. Its non linear dependence on the electric field $E = E_{1\rm n}$ in the dielectric film---and then on the electric potential $v = E\, d$ applied to the dielectric film, $v$ being the potential on a point of $\rm S_1$ minus that on the neighbouring point of $\rm S_{12}$---will be simply modeled using a threshold value $E_0 = v_0/d$ and a high slope $\gamma_0$ for $E > E_0$ ($\gamma_0 \gg \gamma$). For the sake of simplicity in the notations, we will use the same threshold value and slope in the region $E < 0$, i.e.,
\begin{align}
j_{\rm E} &= 0 \quad{\rm if}\quad -E_0 < E < E_0 \nonumber\\
j_{\rm E} &= \gamma_0 (E - E_0) \quad{\rm if}\quad E > E_0 \nonumber\\
j_{\rm E} &= \gamma_0 (E + E_0) \quad{\rm if}\quad E < -E_0 \label{jE(v)}
\end{align}
(Fig.~\ref{E current}).\cite{Asymmetry} Denoting $j = j_{2\rm n} = \gamma\,E_{2\rm n}$, the current in the solution (at $\rm S_{12}$), Eq.~(\ref{Interface}) may thus be written as
\begin{eqnarray}
j = j_{\rm C} + j_{\rm E}.\label{j}
\end{eqnarray}

\begin{figure}[htbp]
\begin{center}
\includegraphics[width=5cm]{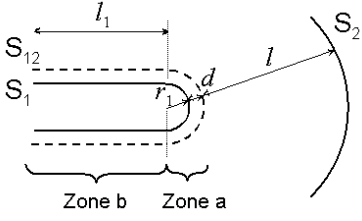}
\end{center}
\caption{Geometrical model of the nanoelectrode surface $\rm S_1$: a hemisphere (zone $\rm a$) and a cylinder (zone $\rm b$).} \label{S1 geometry}
\end{figure}

At a given time $t$, all the points of $\rm S_1$ (of $\rm S_2$, respectively) have practically the same potential---because the electrodes are made of metal---and we denote $V(t)$ the potential on $\rm S_1$ minus that on $\rm S_2$. On the contrary, the surface $\rm S_{12}$ is not equipotential because the thickness $d$ of the dielectric film is constant but the electric field (in this film) varies, being higher at the apex of the nanoelectrode (where the curvature radius of $\rm S_1$ is very small). Then, at a given time $t$, the potential value $v$ varies with the position on $\rm S_1$ (or on $\rm S_{12}$). In the following, the nanoelectrode surface $\rm S_1$ will be simply modeled as (i) a hemisphere of radius $r_1$ at the apex, denoted zone $\rm a$, and (ii) a cylinder (of the same radius) of length $l_1$, denoted zone $\rm b$ (Fig.~\ref{S1 geometry}). 
As a first approximation, we assume that the electric potential field in zone $\rm a$ (except near the junction with the cylinder of zone $\rm b$) is that produced by a spherical electrode and that in zone $\rm b$ (except near the two ends of the cylinder) is that produced by a (infinite) cylindrical electrode. With the help of Maxwell's equations, we then obtain the values of $E_{1\rm n}$ and $E_{2\rm n}$ in zones $\rm a$ and $\rm b$ 
\begin{align}
E_{1\rm n} &= \beta_1 v \nonumber\\
E_{2\rm n} &= \beta_2 (V - v), \label{E1n-E2n}
\end{align}
the values of $\beta_1$, $\beta_2$, and $v$ being different in zones $\rm a$ and $\rm b$, and distinguished with the respective subscripts $\rm a$ and $\rm b$, i.e.,
\begin{align}
\beta_{\rm 1a} &= \frac{r_1}{(r_1 + d)d} \approx \frac{1}{d} \nonumber\\
\beta_{\rm 2a} &= \frac{r_1 + d + l}{(r_1 + d)l} \approx \frac{1}{r_1} \label{beta1a-beta2a}
\end{align}
in zone $\rm a$, and
\begin{align}
\beta_{\rm 1b} &= \frac{1}{(r_1 + d)\log \frac{r_1 + d}{r_1}} \approx \frac{1}{d} \nonumber\\
\beta_{\rm 2b} &= \frac{1}{(r_1 + d)\log \frac{r_1 + d + l}{r_1 + d}} 
\approx \frac{1}{r_1\log \frac{l}{r_1}} \label{beta1b-beta2b}
\end{align}
in zone $\rm b$ (see Appendix \ref{Sph Cyl}; $\log$ refers to the natural logarithm; the approximations hold if $d \ll r_1 \ll l$).

Equation~(\ref{Interface}) may then be written as
\begin{eqnarray*}
\gamma\beta_2 (V - v) = -\varepsilon\frac{d}{d t}(\beta_2 (V - v)-\beta_1 v) + j_{\rm E},
\end{eqnarray*}
i.e.,
\begin{eqnarray}
V - v = \tau(\frac{dv}{dt} - \frac{1}{\alpha}\frac{dV}{dt}) + \frac{j_{\rm E}}{\gamma\beta_2},\label{vV}
\end{eqnarray}
where $\tau = \frac{\varepsilon}{\gamma} \alpha$ and $\alpha = 1 + \frac{\beta_1}{\beta_2}$, which, according to Eq.~(\ref{jE(v)}) (with $E = v/d$ and $E_0 = v_0/d$), leads to
\begin{align}
v + \tau \frac{dv}{dt} 
&= V + \frac{\tau}{\alpha} \frac{dV}{dt}\quad{\rm if}\quad -v_0 < v < v_0,\label{-v0v0}\\
v + \tau' \frac{dv}{dt} 
&= V_+ + \frac{\tau}{\alpha} \frac{dV_+}{dt}\quad{\rm if}\quad v > v_0,\label{v0}\\
v + \tau' \frac{dv}{dt} 
&= V_- + \frac{\tau}{\alpha} \frac{dV_-}{dt}\quad{\rm if}\quad v < -v_0,\label{-v0} 
\end{align}
where $\tau' = \frac{\tau}{1 + \tilde\gamma \alpha'}$, $\tilde\gamma = \frac{\gamma_0}{\gamma}$, $\alpha' = \frac{1}{\beta_2 d} = \frac{\alpha - 1}{\beta_1 d}$, $V_+ = \frac{V + \tilde\gamma \alpha' v_0}{1 + \tilde\gamma \alpha'}$, and $V_- = \frac{V - \tilde\gamma \alpha' v_0}{1 + \tilde\gamma \alpha'}$. Note that $\alpha$, $\alpha'$, $\tau$, and $\tau'$ have different values in zones $\rm a$ and $\rm b$, which will be distinguished with the respective subscripts $\rm a$ and $\rm b$. For a given applied potential $t \rightarrow V(t)$, Eq.~(\ref{vV}) or Eqs.~(\ref{-v0v0})--(\ref{-v0}) determine $v$ as a function of time [i.e., $v_{\rm a}(t)$ using $\alpha_{\rm a}$, $\alpha'_{\rm a}$, $\tau_{\rm a}$, and $\tau'_{\rm a}$, and $v_{\rm b}(t)$ using $\alpha_{\rm b}$, $\alpha'_{\rm b}$, $\tau_{\rm b}$, and $\tau'_{\rm b}$].

Since $j = \gamma\,E_{2\rm n} = \gamma \beta_2 (V - v)$, the current intensity (in the solution, in zone $\rm a$ or in zone $\rm b$) is
\begin{eqnarray}
I = \gamma \beta_2 a (V - v) = \frac{\gamma a}{\alpha' d} (V - v), \label{I}
\end{eqnarray}
$a$ being the corresponding area of $\rm S_{12}$, i.e., $a_{\rm a} = 2\pi(r_1 + d)^2 \approx 2\pi r_1^2$ for zone $\rm a$ and $a_{\rm b} = 2\pi(r_1 + d)l_1 \approx 2\pi r_1 l_1$ for zone $\rm b$. In fact, the total intensity in the electric circuit is the sum of the contributions of zone $\rm a$ and zone $\rm b$:
\begin{eqnarray}
I = I_{\rm a} + I_{\rm b}. \label{Itotal}
\end{eqnarray}
Similarly, the electrolysis current intensity and the charging current intensity (in zone $\rm a$ or in zone $\rm b$) are, respectively, $I_{\rm E} = j_{\rm E}\; a$ and $I_{\rm C} = j_{\rm C}\; a$; thus
\begin{align}
I_{\rm E} &= 0 \quad{\rm if}\quad -v_0 < v < v_0 \nonumber\\
I_{\rm E} &= \frac{\gamma a}{\alpha' d}\,\tilde\gamma \alpha'(v - v_0) \quad{\rm if}\quad v > v_0 \nonumber\\
I_{\rm E} &= \frac{\gamma a}{\alpha' d}\,\tilde\gamma \alpha'(v + v_0) \quad{\rm if}\quad v < -v_0, \label{IE}
\end{align}
and
\begin{eqnarray}
I = I_{\rm C} + I_{\rm E} \label{ICE}
\end{eqnarray}
(in zone $\rm a$ or in zone $\rm b$).

\section{A simple example: potential of rectangular shape}

\begin{figure}[htbp]
\begin{center}
\includegraphics[width=5cm]{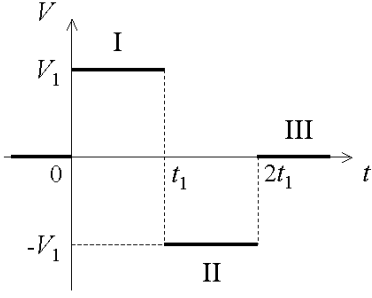}
\end{center}
\caption{Applied potential $t \rightarrow V(t)$ of rectangular shape.} \label{V(t)}
\end{figure}

Let us show the consequences of the preceding model with a simple example. The simplest case is the application of a constant electric potential $V_1$ (between the two electrodes) during a finite time $t_1$ and is treated below as phase I (Sec.~\ref{IIIA}). The nanoelectrode is thus anode (cathode, respectively) during the time $t_1$ if $V_1 > 0$ ($V_1 < 0$, respectively). In order to treat both cases (anode and cathode) and to study the possible occurrence of both oxidation and reduction reactions (in zone a and in zone b), we will consider the simple case of a potential of rectangular shape in which the preceding phase I is followed by a second phase (phase II) with a constant opposite potential $-V_1$ during the same time $t_1$, after which no potential is applied (phase III) (Fig.~\ref{V(t)})
\begin{align*}
V(t) &= 0 \quad{\rm if}\quad t < 0 \\
V(t) &= V_1 \quad{\rm if}\quad 0 < t < t_1 \quad{\rm (phase\;I)}\\
V(t) &= -V_1 \quad{\rm if}\quad t_1 < t < 2 t_1 \quad{\rm (phase\;II)}\\
V(t) &= 0 \quad{\rm if}\quad t > 2 t_1 \quad{\rm (phase\;III)}.
\end{align*}

We here consider $V_1 > 0$, but the case $V_1 < 0$ is exactly similar. In Sec.~\ref{IIIA}, we will see that the oxidation reaction may occur or not during phase I, in zone a and in zone b, depending on the values of $V_1$ and $t_1$. This will explain the nanolocalization of the oxidation reaction (when this reaction only occurs at the apex of the nanoelectrode, i.e., in zone a but not in zone b). Similarly, Sec.~\ref{IIIB} concerns the possible occurrence of the reduction reaction during phase II and its nanolocalization at the apex of the electrode.

\subsection{Phase I} \label{IIIA}

\subsubsection{Before the electrolysis reaction}

\begin{figure}[htbp]
\begin{center}
\includegraphics[width=8.5cm]{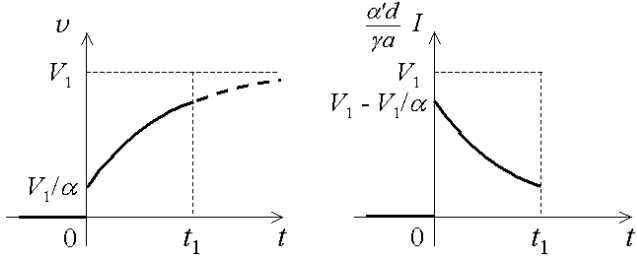}
\end{center}
\caption{Solution $v$ of Eq.~(\ref{-v0v0}) during phase I (left) and corresponding current intensity $I$ (right).} \label{v-I}
\end{figure}

During phase I, let us consider the solution $v$ of Eq.~(\ref{vV}) or (\ref{-v0v0}) (in zone $\rm a$ or in zone $\rm b$), assuming that $v(t)$ remains lower than $v_0$. At $t = 0$, the discontinuity jump $V(0^+) - V(0^-) = V_1$ produces the term $V_1 \delta_0(t)$ in $\frac{dV}{dt}$ ($\delta_0$ being the Dirac measure at 0), then, according to Eq.~(\ref{vV}), the term $\frac{V_1}{\alpha} \delta_0(t)$ in $\frac{dv}{dt}$, and then the discontinuity jump $v(0^+) - v(0^-) = \frac{V_1}{\alpha}$ for $v$ at 0. If $V_1 < \alpha\, v_0$ (which corresponds to our usual experimental conditions\cite{Values}), the solution is then
\begin{align}
v(t) &= 0 \quad{\rm if}\quad t < 0 \nonumber\\
v(t) &= V_1 - (V_1 - \frac{V_1}{\alpha})\,e^{-t/\tau} \quad{\rm if}\quad t > 0 \label{vI}
\end{align}
represented in Fig.~\ref{v-I}, with the corresponding current intensity given by Eq.~(\ref{I}).

\begin{figure}[htbp]
\begin{center}
\includegraphics[width=4cm]{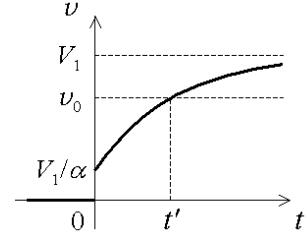}
\end{center}
\caption{Solution $v$ of Eq.~(\ref{-v0v0}) during phase I. At the time $t'$, $v(t)$ reaches the value $v_0$ (if the duration of phase I is large enough).} \label{v-t'}
\end{figure}

Clearly, $v(t)$ is always lower than $V_1$ and, if $V_1 < v_0$, $v(t)$ will always remain lower than $v_0$. In the following, we suppose $v_0 < V_1 < \alpha\, v_0$, so that $v(t)$ will reach the value $v_0$ at the time
\begin{eqnarray}
t' = \tau \log\frac{(1 - \frac{1}{\alpha})V_1}{V_1 - v_0} \label{t'}
\end{eqnarray}
(if the duration of phase I is large enough; Fig.~\ref{v-t'}).
Note that if $V_1 > \alpha\, v_0$, then $v(0^+) = \frac{V_1}{\alpha} > v_0$, so that in this case $t' = 0$ [and the solution of Eq.~(\ref{v0}) shows that $v(t) > v_0$ for $0 < t < t_1$].

\begin{figure}[htbp]
\begin{center}
\includegraphics[width=6.5cm]{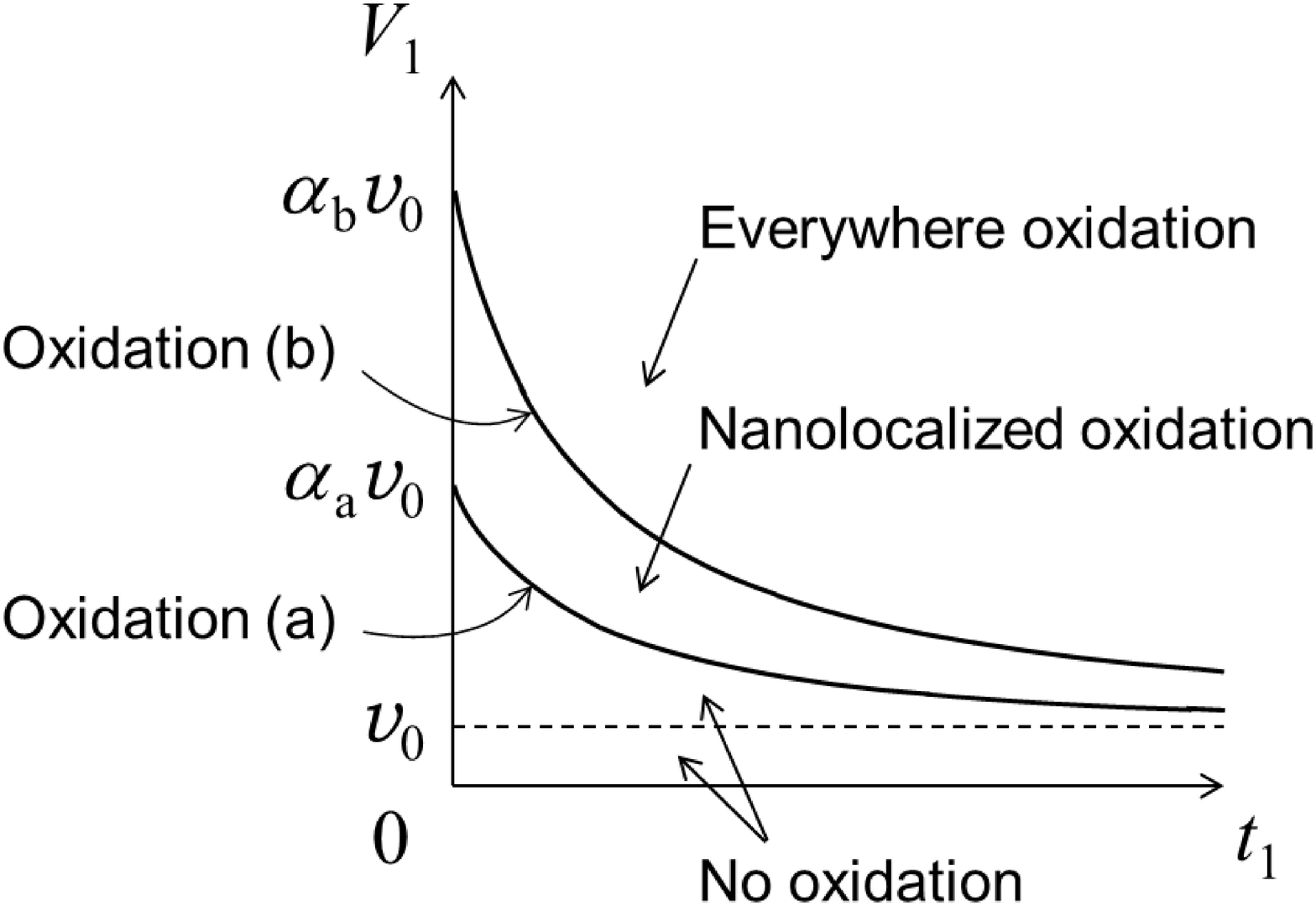}
\end{center}
\caption{During phase I, in the plane $(t_1,V_1)$, the two oxidation curves $t_1 = t'_{\rm a}(V_1)$ in zone a and $t_1 = t'_{\rm b}(V_1)$ in zone b delimit the three regions: ``no oxidation'' (lower region), ``nanolocalized oxidation'' (between the two curves), and ``everywhere oxidation'' (upper region).} \label{t'a-t'b}
\end{figure}

Consider now the relative position of $t_1$ and $t'$. If $t' > t_1$, $v(t)$ will not reach the value $v_0$ during phase I (i.e., for $0 < t < t_1$), and if $t' < t_1$, $v(t)$ will reach the value $v_0$ during phase I, at the time $t'$. In this last case, according to Eq.~(\ref{jE(v)}), there will be an electrolysis current $j_{\rm E} > 0$, i.e., the oxidation reaction of electrolysis will occur, for $t' < t < t_1$. Let us represent, in the plane $(t_1,V_1)$, the two ``oxidation'' curves $t_1 = t'_{\rm a}(V_1)$ in zone a (given by Eq.~(\ref{t'}) with $\tau_{\rm a}$ and $\alpha_{\rm a}$) and $t_1 = t'_{\rm b}(V_1)$ in zone b [Eq.~(\ref{t'}) with $\tau_{\rm b}$ and $\alpha_{\rm b}$] (Fig.~\ref{t'a-t'b}).
In the region $t_1 < t'_{\rm a}(V_1)$, there is no oxidation reaction (neither in zone ${\rm a}$ nor in zone ${\rm b}$). In the region $t_1 > t'_{\rm b}(V_1)$, the oxidation reaction occurs
everywhere on the nanoelectrode (in zone ${\rm a}$, after $t'_{\rm a}$, and in zone ${\rm b}$, after $t'_{\rm b}$). {\it In the region $t'_{\rm a}(V_1) < t_1 < t'_{\rm b}(V_1)$, the oxidation reaction is nanolocalized at the apex of the nanoelectrode} (it occurs in zone ${\rm a}$, after $t'_{\rm a}$, but not in zone ${\rm b}$).

\subsubsection{During the electrolysis reaction} \label{IIIA2}

\begin{figure}[htbp]
\begin{center}
\includegraphics[width=8.5cm]{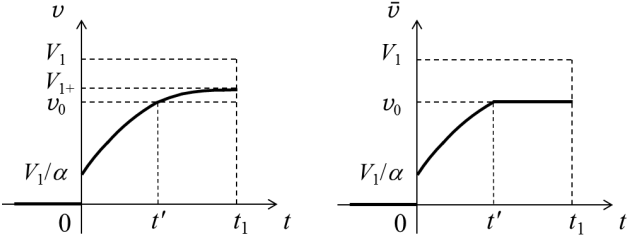}
\end{center}
\caption{Solution $v$ during phase I (i.e., Eq.~(\ref{vI}) for $t < t'$ and Eq.~(\ref{vIE}) for $t' < t < t_1$) for a finite value of $\tilde\gamma$ (left) and for $\tilde\gamma = +\infty$ (right). Case $t' < t_1$.} \label{v(I)}
\end{figure}

\begin{figure}[htbp]
\begin{center}
\includegraphics[width=8.5cm]{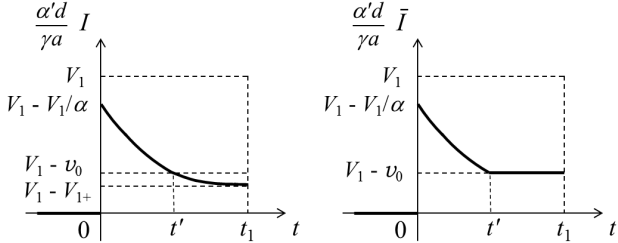}
\end{center}
\caption{Current intensity $I$ during phase I for $\tilde\gamma < +\infty$ (left) and for $\tilde\gamma = +\infty$ (right). Case $t' < t_1$.} \label{I(I)}
\end{figure}

\begin{figure}[htbp]
\begin{center}
\includegraphics[width=8.5cm]{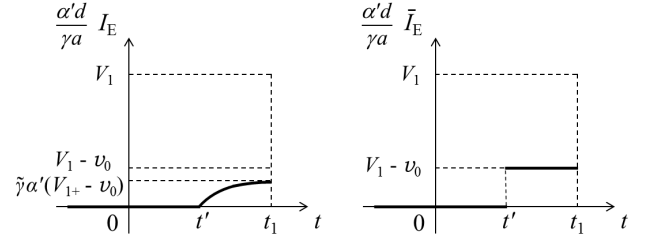}
\end{center}
\caption{Electrolysis current intensity $I_{\rm E}$ during phase I for $\tilde\gamma < +\infty$ (left) and for $\tilde\gamma = +\infty$ (right). Case $t' < t_1$.} \label{IE(I)}
\end{figure}

If $t' < t_1$ (in zone ${\rm a}$ or in zone ${\rm b}$), $v(t)$ reaches the value $v_0$ at $t'$ and, for $t > t'$, $v$ is the solution of the new Eq.~(\ref{v0}), i.e.,
\begin{eqnarray}
v(t) = V_{1+} - (V_{1+} - v_0)\,e^{-(t - t')/\tau'} \quad{\rm if}\quad t > t', \label{vIE}
\end{eqnarray}
where $V_{1+} = \frac{V_1 + \tilde\gamma \alpha' v_0}{1 + \tilde\gamma \alpha'}$, represented in Fig.~\ref{v(I)}. Note that, for any finite value of $\tilde\gamma$, $dv/dt$ is continuous at $t'$ and $\frac{dv}{dt}(t') = (V_1 - v_0)/\tau$.
Since $\gamma_0 \gg \gamma$, we may consider that $\tilde\gamma$ tends to $+\infty$ and the corresponding limit value of any quantity $X$ will be denoted $\bar X$. Thus, $v(t)$ tends to $\bar v(t) = v_0$ for $t' < t < t_1$ (since $V_{1+}$ tends to $v_0$; see Fig.~\ref{v(I)}). According to Eq.~(\ref{I}), the current intensity $I$ and its limit value $\bar I$ are then represented in Fig.~\ref{I(I)}.
Similarly, Eq.~(\ref{IE}) gives the electrolysis current intensity $I_{\rm E}$ and its limit value $\bar I_{\rm E}$ (Fig.~\ref{IE(I)}). 
Note that $I$ and $I_{\rm E}$ have the same asymptotic value $\frac{\gamma a}{\alpha' d} (V_1 - V_{1+}) =\frac{\gamma a}{\alpha' d}\,\tilde\gamma \alpha'(V_{1+} - v_0) = \frac{\gamma a}{\alpha' d} \frac{\tilde\gamma \alpha'}{1 + \tilde\gamma \alpha'} (V_1 - v_0)$ and that $\bar I = \bar I_{\rm E} = \frac{\gamma a}{\alpha' d} (V_1 - v_0)$ for $t' < t < t_1$. Thus, while before the electrolysis reaction ($0 < t < t'$), the current intensity is equal to the charging current intensity ($I = I_{\rm C}, I_{\rm E} = 0$), {\it during the electrolysis reaction ($t' < t < t_1$), owing to $\gamma_0 \gg \gamma$, the (limit) current intensity is equal to the (limit) electrolysis current intensity ($\bar I = \bar I_{\rm E}, \bar I_{\rm C} = 0$)}. This is a consequence of $\bar v = v_0 =$ constant for $t' < t < t_1$, which implies that $E_{1\rm n}$ and $E_{2\rm n}$ are constant [from Eq.~(\ref{E1n-E2n})], and then $\bar j_{\rm C} = -\varepsilon\frac{\partial}{\partial t}(E_{2\rm n}-E_{1\rm n}) = 0$.

From Eqs.~(\ref{IE}) and (\ref{vIE}), we then obtain the electric charge given by the nanoelectrode to the solution for the electrolysis reaction (in zone a or zone b)
\begin{align}
Q' &= \int_{t'}^{t_1} I_{\rm E}(t) dt \nonumber\\ 
&= \frac{\gamma a}{\alpha' d}\,\frac{\tilde\gamma \alpha'}{1 + \tilde\gamma \alpha'}\,(V_1 - v_0)(T' - \tau'(1 - e^{-T'/\tau'})), \label{Q'}
\end{align}
where $T' = t_1 - t'$ is the electrolysis duration, and its limit value (for $\tilde\gamma = +\infty$)
\begin{eqnarray}
\bar Q' = \frac{\gamma a}{\alpha' d}\,(V_1 - v_0)\,T'. \label{bar Q'}
\end{eqnarray}
Note that this charge is due to the electrons ---tunneling through the dielectric film, from the solution to the nanoelectrode--- produced by the electrolysis oxidation reaction
\begin{eqnarray*}
\rm 2\;H_2O \rightarrow O_2 + 4\;H^+ + 4\;e^-, 
\end{eqnarray*}
which gives the number of produced $\rm O_2$ molecules
\begin{eqnarray}
n_{\rm O_2} = \frac{Q'}{4\,q_{\rm e}} \label{nO2}
\end{eqnarray}
($q_{\rm e}$ the elementary charge).

\subsection{Phase II} \label{IIIB}

\subsubsection{Before the electrolysis reaction}

First note that the following results are based on the simple assumption of a unique threshold value $v_0$ (and slope $\gamma_0$).\cite{Asymmetry} As previously mentioned, the following study relates to either zone a or zone b. Because of Eq.~(\ref{vV}), the discontinuity jump $V(t_1^+) - V(t_1^-) = -2 V_1$ produces the discontinuity jump $v(t_1^+) - v(t_1^-) = -2 \frac{V_1}{\alpha}$ (as it occurred at $t = 0$). Since $-v_0 < v(t_1^+) < v_0$ ($v(t_1^+) > -v_0$ because $v(t_1^-) > \frac{V_1}{\alpha}$; $v(t_1^+) < v_0$ because $v(t_1^-)$ is either $< v_0$ or $< V_{1+} \approx v_0$ owing to the high value of $\tilde\gamma$), the solution of Eq.~(\ref{-v0v0}) gives
\begin{eqnarray}
v(t) = -V_1 + (V_1 + v(t_1^+))\,e^{-(t - t_1)/\tau} \;\,{\rm if}\;\, t > t_1 \label{vII}
\end{eqnarray}
(as long as $v(t) > -v_0$). In the case $t_1 < t'$, we know that, during phase I, $v(0^+) = \frac{V_1}{\alpha}$ and $v(t)$ increases but does not reach the value $v_0$ for $0 < t < t_1$. If $v(t_1^+)$ were equal to $-\frac{V_1}{\alpha}$, we would have the same situation (with the opposite sign for $v$) during phase II, so that $v(t)$ (now decreasing) would not reach the value $-v_0$ for $t_1 < t < 2t_1$. Since $v(t_1^+) > -\frac{V_1}{\alpha}$ (because $v(t_1^-) > \frac{V_1}{\alpha}$), this implies that $v(t)$ will not reach the value $-v_0$ during phase II. In other words, {\it if there is no electrolysis reaction (oxidation) during phase I, there will be no electrolysis reaction (reduction) during phase II}.

We then consider the case $t' < t_1$. If the duration of phase II is large enough, $v(t)$ will reach the value $-v_0$ at the time $t_1 + t''$, with [from Eq.~(\ref{vII})]
\begin{align}
t'' &= \tau \log\frac{V_1 + v(t_1^+)}{V_1 - v_0} \nonumber\\ 
&= \tau \log\frac{(1-\frac{2}{\alpha})V_1 + v_0 + \frac{V_1 - v_0}{1 + \tilde\gamma \alpha'}(1 - e^{-(t_1 - t')/\tau'})}{V_1 - v_0} \label{t''}
\end{align}
[using $v(t_1^-)$ given by Eq.~(\ref{vIE})] and its limit value (for $\tilde\gamma = +\infty$)
\begin{eqnarray}
\bar t'' =  \tau \log\frac{(1-\frac{2}{\alpha})V_1 + v_0}{V_1 - v_0}. \label{bar t''}
\end{eqnarray}

Clearly, $t'' > t'$ because $v(t_1^+) > -\frac{V_1}{\alpha}$. Thus, if $t'' < t_1$, {\it the electrolysis duration (reduction) $T'' = t_1 - t''$ during phase II is lower than the electrolysis duration (oxidation) $T' = t_1 - t'$ during phase I}.
 
Strictly speaking, $t''$ is a function of $(t_1, V_1)$ [Eq.~(\ref{t''}) with $t'$ given by Eq.~(\ref{t'})], but its dependence on $t_1$ is very low and, for large values of $\tilde\gamma$, $t'' \approx \bar t''$ which only depends on $V_1$. Let us represent, in the plane $(t_1,V_1)$, the two ``reduction'' curves $t_1 = t''_{\rm a}(t_1, V_1)$ in zone a [given by Eq.~(\ref{t''}) with $\tau_{\rm a}$, $\alpha_{\rm a}$, $\tau'_{\rm a}$, $\alpha'_{\rm a}$, and $t'_{\rm a}(V_1)$] and $t_1 = t''_{\rm b}(t_1, V_1)$ in zone b [Eq.~(\ref{t''}) with $\tau_{\rm b}$, $\alpha_{\rm b}$, $\tau'_{\rm b}$, $\alpha'_{\rm b}$, and $t'_{\rm b}(V_1)$]. Thus, in the region $t_1 < t''_{\rm a}(t_1, V_1)$, there is no reduction reaction of electrolysis during phase II (neither in zone ${\rm a}$ nor in zone ${\rm b}$). In the region $t_1 > t''_{\rm b}(t_1, V_1)$, the reduction reaction occurs
everywhere on the nanoelectrode (in zone ${\rm a}$, after $t_1 + t''_{\rm a}$, and in zone ${\rm b}$, after $t_1 + t''_{\rm b}$). {\it In the region $t''_{\rm a}(t_1, V_1) < t_1 < t''_{\rm b}(t_1, V_1)$, the reduction reaction is nanolocalized at the apex of the nanoelectrode} (it occurs in zone ${\rm a}$, after $t_1 + t''_{\rm a}$, but not in zone ${\rm b}$).

\begin{figure}[htbp]
\begin{center}
\includegraphics[width=8cm]{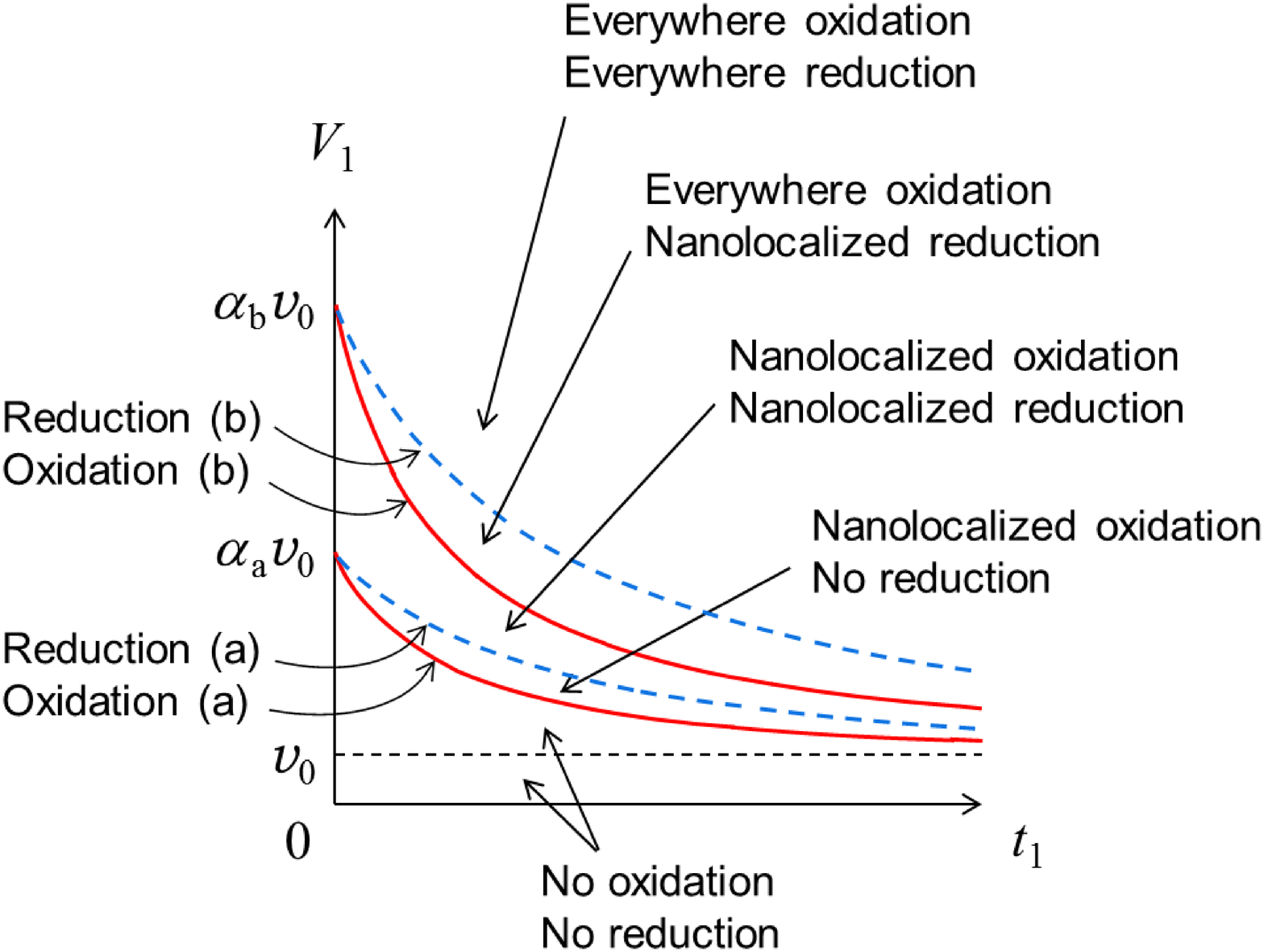}
\end{center}
\caption{In the plane $(t_1,V_1)$, the four curves $t_1 = t'_{\rm a}$ (oxidation in zone a), $t_1 = t''_{\rm a}$ (reduction in zone a), $t_1 = t'_{\rm b}$ (oxidation in zone b), and $t_1 = t''_{\rm b}$ (reduction in zone b) delimit the five regions: ``no oxidation, no reduction'' (below $t_1 = t'_{\rm a}$), ``nanolocalized oxidation, no reduction'' (between $t_1 = t'_{\rm a}$ and $t_1 = t''_{\rm a}$), ``nanolocalized oxidation, nanolocalized reduction'' (between $t_1 = t''_{\rm a}$ and $t_1 = t'_{\rm b}$), ``everywhere oxidation, nanolocalized reduction'' (between $t_1 = t'_{\rm b}$ and $t_1 = t''_{\rm b}$), and ``everywhere oxidation, everywhere reduction'' (above $t_1 = t''_{\rm b}$).} \label{t'ab-t''ab}
\end{figure}

It may be shown that $t''_{\rm a} < t'_{\rm b}$ in general (e.g., if $\alpha_{\rm b} > 2 \alpha_{\rm a}$, $\alpha_{\rm a} \gg 1$, and $\tilde\gamma > 1$) for any $(t_1,V_1)$ such that $v_0 < V_1 < \alpha_{\rm a} v_0$ and $t_1 > t'_{\rm a}$. We thus have five regions in the plane $(t_1,V_1)$ (see Fig.~\ref{t'ab-t''ab}).
Below the curve $t_1 = t'_{\rm a}$, there is no oxidation reaction and no reduction reaction (neither in zone ${\rm a}$ nor in zone ${\rm b}$). {\it Between the two curves $t_1 = t'_{\rm a}$ and $t_1 = t''_{\rm a}$, the oxidation reaction is nanolocalized at the apex of the nanoelectrode} (it occurs in zone ${\rm a}$ but not in zone ${\rm b}$) {\it and there is no reduction reaction} (neither in zone ${\rm a}$ nor in zone ${\rm b}$). {\it Between the two curves $t_1 = t''_{\rm a}$ and $t_1 = t'_{\rm b}$, the oxidation reaction and the reduction reaction are nanolocalized at the apex of the nanoelectrode} (they occur in zone ${\rm a}$ but not in zone ${\rm b}$). {\it Between the two curves $t_1 = t'_{\rm b}$ and $t_1 = t''_{\rm b}$, the oxidation reaction occurs everywhere on the nanoelectrode} (in zone ${\rm a}$ and in zone ${\rm b}$) {\it and the reduction reaction is nanolocalized at the apex of the nanoelectrode} (it occurs in zone ${\rm a}$ but not in zone ${\rm b}$). Above the curve $t_1 = t''_{\rm b}$, the oxidation reaction and the reduction reaction occur everywhere on the nanoelectrode (in zone ${\rm a}$ and in zone ${\rm b}$).

\subsubsection{During the electrolysis reaction} \label{IIIB2}

\begin{figure}[htbp]
\begin{center}
\includegraphics[width=7.2cm]{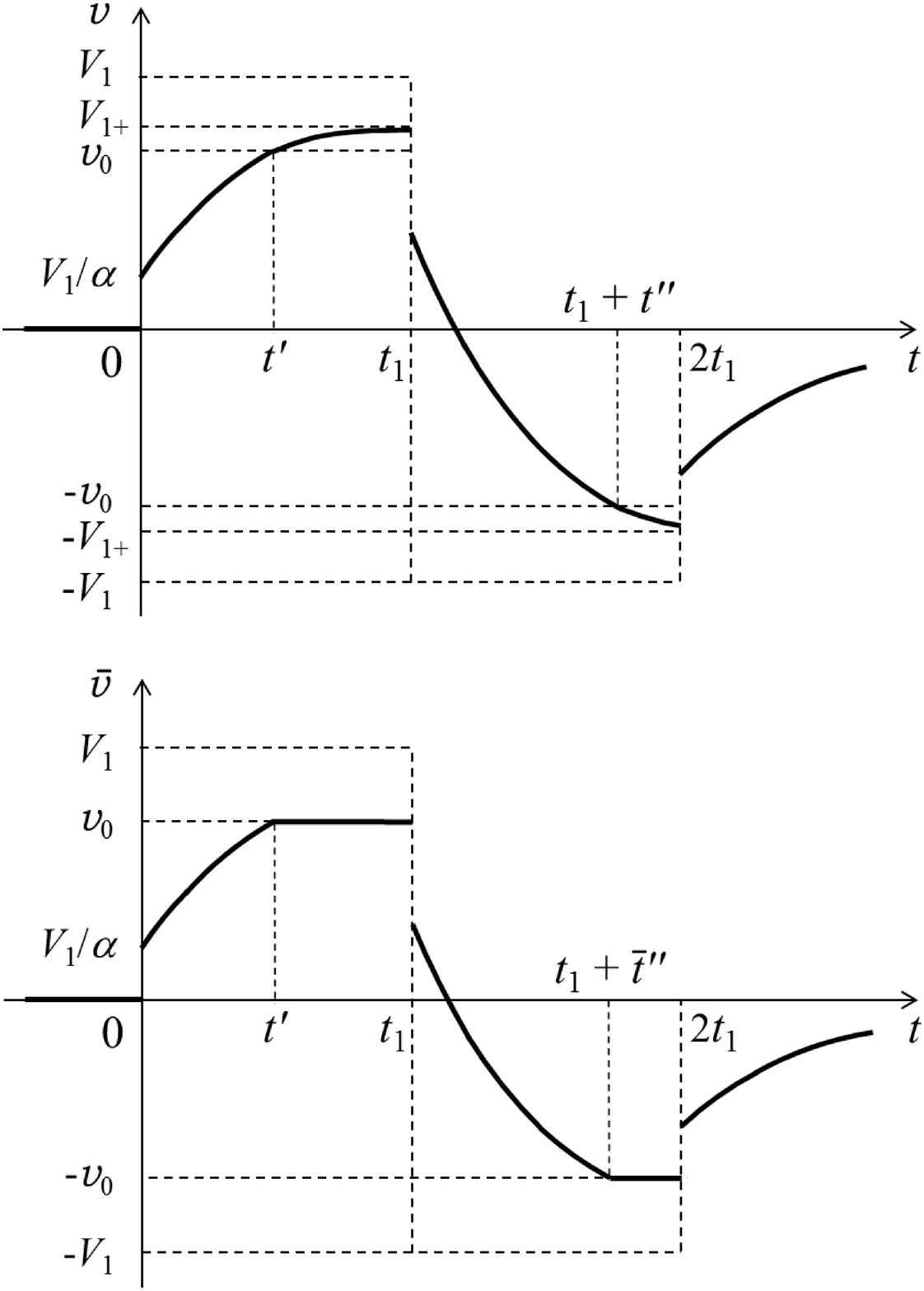}
\end{center}
\caption{Solution $v$ during phase II (i.e., Eq.~(\ref{vII}) for $t_1 < t < t_1 + t''$ and Eq.~(\ref{vIIE}) for $t_1 + t'' < t < 2t_1$) for $\tilde\gamma < +\infty$ (top) and for $\tilde\gamma = +\infty$ (bottom). Case $t'' < t_1$. Phase I and phase III are also represented.} \label{v(II)}
\end{figure}

\begin{figure}[htbp]
\begin{center}
\includegraphics[width=8.6cm]{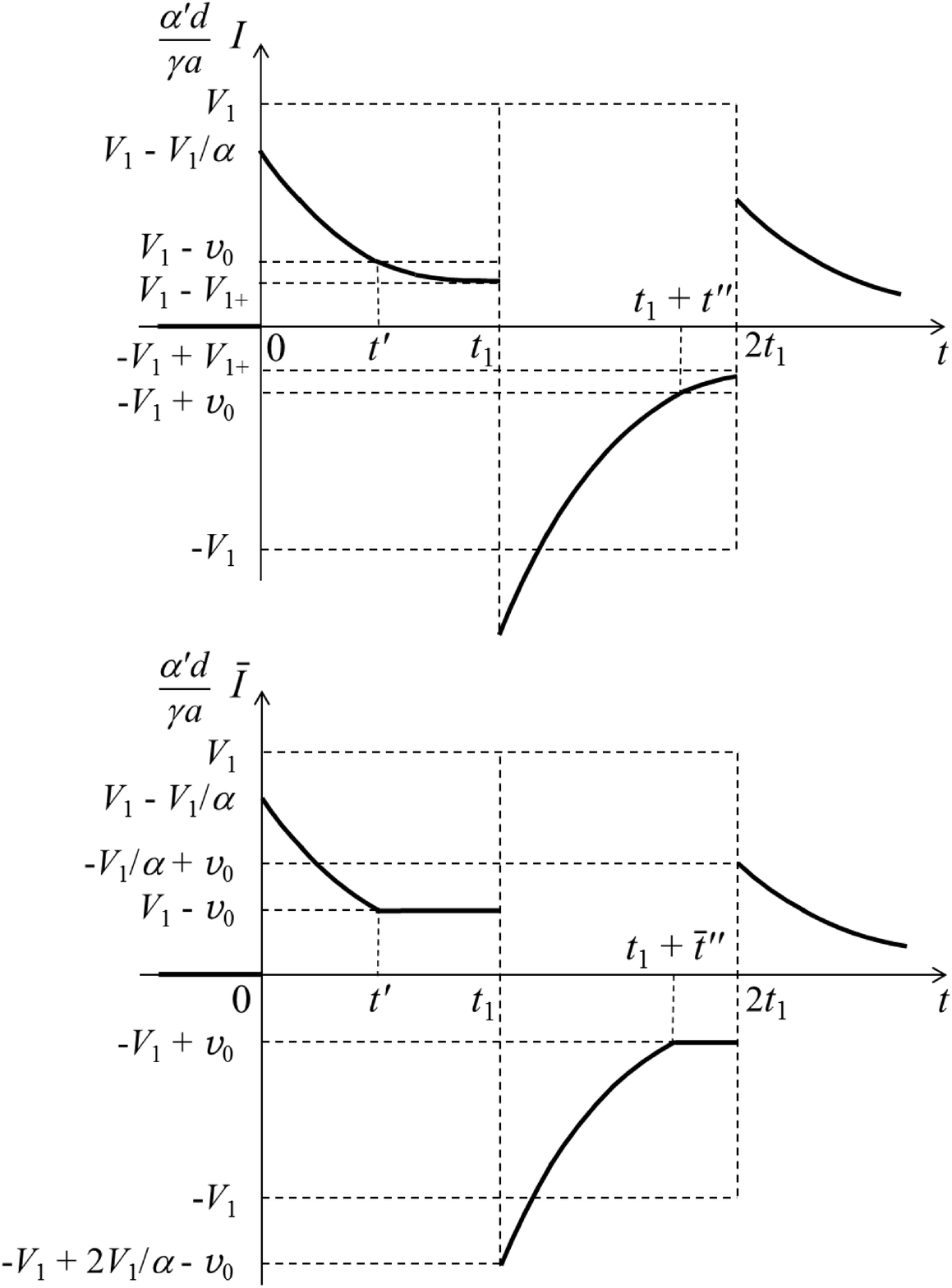}
\end{center}
\caption{Current intensity $I$ during phase II for $\tilde\gamma < +\infty$ (top) and for $\tilde\gamma = +\infty$ (bottom). Case $t'' < t_1$. Phase I and phase III are also represented.} \label{I(II)}
\end{figure}

\begin{figure}[htbp]
\begin{center}
\includegraphics[width=8.6cm]{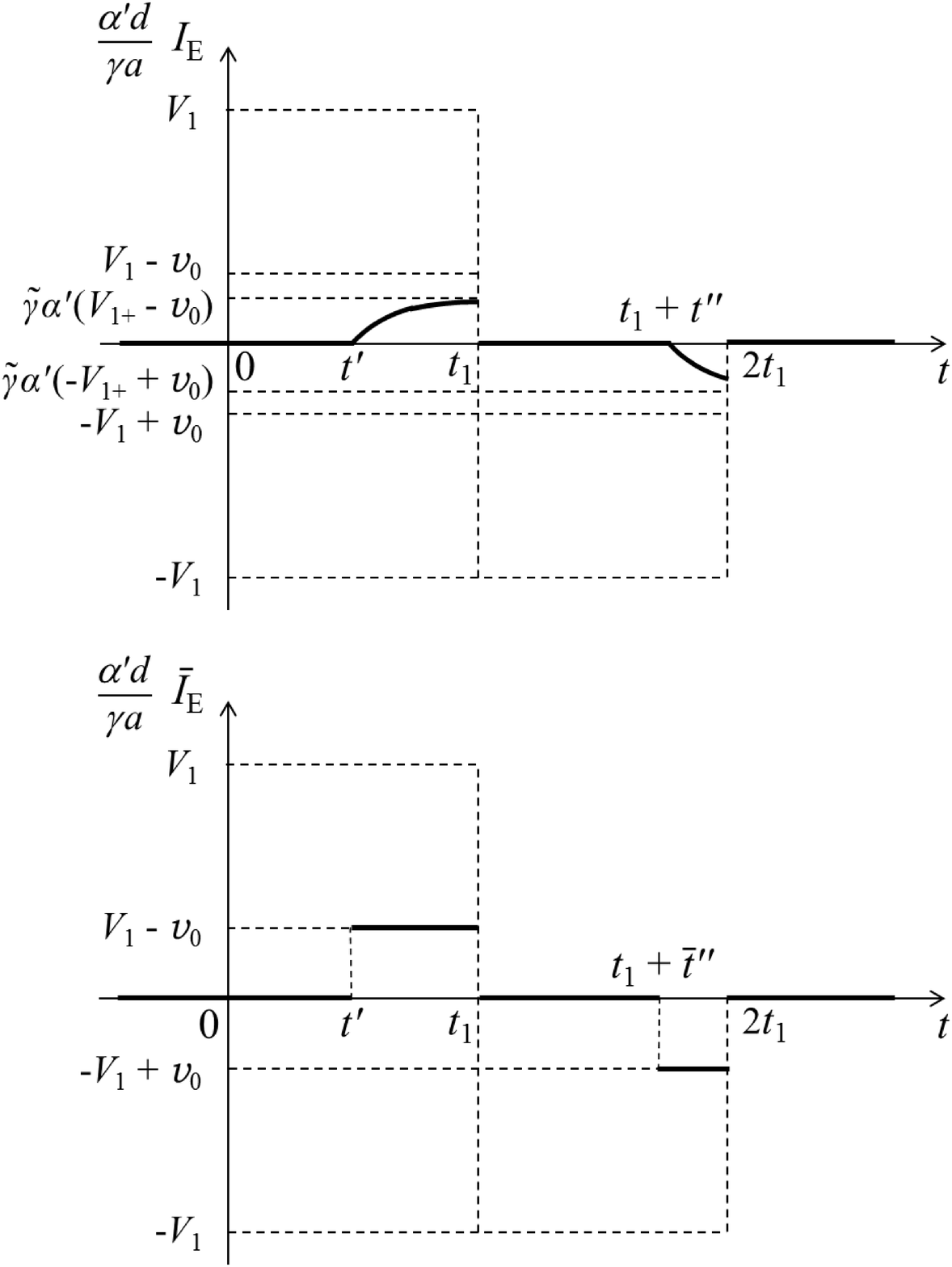}
\end{center}
\caption{Electrolysis current intensity $I_{\rm E}$ during phase II for $\tilde\gamma < +\infty$ (top) and for $\tilde\gamma = +\infty$ (bottom). Case $t'' < t_1$. Phase I and phase III are also represented.} \label{IE(II)}
\end{figure}

If $t'' < t_1$ (in zone ${\rm a}$ or in zone ${\rm b}$), $v(t)$ reaches the value $-v_0$ at $t_1 + t''$ and, for $t > t_1 + t''$, $v$ is the solution of the new Eq.~(\ref{-v0}), i.e.,
\begin{eqnarray}
v(t) = -V_{1+} + (V_{1+} - v_0)\,e^{-(t - t_1 - t'')/\tau'}\nonumber\\
{\rm if}\quad t > t_1 + t'', \label{vIIE}
\end{eqnarray}
$v$ and $\bar v$ being represented in Fig.~\ref{v(II)}. Note that, for $\tilde\gamma < +\infty$, $dv/dt$ is continuous at $t_1 + t''$ and $\frac{dv}{dt}(t_1 + t'') = -\frac{dv}{dt}(t') = -(V_1 - v_0)/\tau$.
According to Eq.~(\ref{I}), the current intensity $I$ and its limit value $\bar I$ are then represented in Fig.~\ref{I(II)}.
Similarly, Eq.~(\ref{IE}) gives the electrolysis current intensity $I_{\rm E}$ and its limit value $\bar I_{\rm E}$ (Fig.~\ref{IE(II)}). 
Note that $I$ and $I_{\rm E}$ have the same asymptotic value $\frac{\gamma a}{\alpha' d} (-V_1 + V_{1+}) =\frac{\gamma a}{\alpha' d}\,\tilde\gamma \alpha'(-V_{1+} + v_0) = -\frac{\gamma a}{\alpha' d} \frac{\tilde\gamma \alpha'}{1 + \tilde\gamma \alpha'} (V_1 - v_0)$ and that $\bar I = \bar I_{\rm E} = \frac{\gamma a}{\alpha' d} (-V_1 + v_0)$ for $t_1 + t'' < t < 2t_1$ (these are the opposite values compared to those of phase I). Just as in phase I, while before the electrolysis reaction ($t_1 < t < t_1 + t''$), the current intensity is equal to the charging current intensity ($I = I_{\rm C}, I_{\rm E} = 0$), {\it during the electrolysis reaction ($t_1 + t'' < t < 2t_1$), owing to $\gamma_0 \gg \gamma$, the (limit) current intensity is equal to the (limit) electrolysis current intensity ($\bar I = \bar I_{\rm E}, \bar I_{\rm C} = 0$)}.

From Eqs.~(\ref{IE}) and (\ref{vIIE}), we then obtain the electric charge given by the nanoelectrode to the solution for the electrolysis reaction (in zone a or in zone b)
\begin{align}
Q'' &= \int_{t_1 + t''}^{2t_1} I_{\rm E}(t) dt \nonumber\\ 
&= -\frac{\gamma a}{\alpha' d}\,\frac{\tilde\gamma \alpha'}{1 + \tilde\gamma \alpha'}\,(V_1 - v_0)(T'' - \tau'(1 - e^{-T''/\tau'})), \label{Q''}
\end{align}
where $T'' = t_1 - t''$ is the electrolysis duration, and its limit value (for $\tilde\gamma = +\infty$)
\begin{eqnarray}
\bar Q'' = -\frac{\gamma a}{\alpha' d}\,(V_1 - v_0)\,\bar T'' \label{bar Q''}
\end{eqnarray}
($\bar T'' = t_1 - \bar t''$). Note that this charge is due to the electrons tunneling through the dielectric film, from the nanoelectrode to the solution, and producing the electrolysis reduction reaction
\begin{eqnarray*}
\rm 2\;H^+ + 2\;e^- \rightarrow H_2 
\end{eqnarray*}
which gives the number of produced $\rm H_2$ molecules
\begin{eqnarray}
n_{\rm H_2} = \frac{|Q''|}{2\,q_{\rm e}}. \label{nH2}
\end{eqnarray}
As noted earlier, $T'' < T'$, which, according to Eqs.~(\ref{Q'}) and (\ref{Q''}), implies that $|Q''| < Q'$: the absolute value of {\it the electric charge used for the electrolysis reaction during phase II (in our case, reduction) is lower than that used for the electrolysis reaction during phase I (in our case, oxidation)}. In our case, this means that {\it the number of produced $H_2$ molecules is lower than twice the number of produced $O_2$ molecules}.

\subsection{Phase III}

Because of Eq.~(\ref{vV}), the discontinuity jump $V(2t_1^+) - V(2t_1^-) = V_1$ produces the discontinuity jump $v(2t_1^+) - v(2t_1^-) = \frac{V_1}{\alpha}$. Since $-v_0 < v(2t_1^+) < 0$ ($v(2t_1^+) > -v_0$ because $v(2t_1^-)$ is either $> -v_0$ or $> -V_{1+} \approx -v_0$ owing to the high value of $\tilde\gamma$; $v(2t_1^+) < 0$ is proved in Appendix \ref{vIII<0}), the solution of Eq.~(\ref{-v0v0}) gives
\begin{eqnarray}
v(t) = v(2t_1^+)\,e^{-(t - 2t_1)/\tau} \;\,{\rm if}\;\, t > 2t_1, \label{vIII}
\end{eqnarray}
which has the form represented in phase III of Fig.~\ref{v(II)}. Thus, there is no electrolysis reaction during phase III.

\section{Bubble production} \label{IV}

If $t_1 > t'$ (in zone a or in zone b), the electrolysis oxidation reaction (during phase I, between $t'$ and $t_1$) produces $\rm O_2$ molecules which diffuse in the solution. If the volume mole density $\rho$ of these molecules in the solution exceeds the saturation value $H p_{\rm a}$ ($H$ the Henry's constant for $\rm O_2$ in water, $p_{\rm a}$ the atmospheric pressure) in some region of the solution, at some time, a bubble of $\rm O_2$ will be produced. In the simple case $\tilde\gamma = +\infty$, the flux of $\rm O_2$ moles produced at the surface of the nanoelectrode (more precisely, at the interface $\rm S_{12}$), in zone a or in zone b, is
\begin{eqnarray}
j_{\rm m} = \frac{\bar j_{\rm E}}{4 N_{\rm A} q_{\rm e}} 
= \frac{\gamma}{4 N_{\rm A} q_{\rm e}\, \alpha' d} (V_1 - v_0) \label{O2 flux}
\end{eqnarray}
constant between $t'$ and $t_1$ (see Sec.~\ref{IIIA2} and bottom of Fig.~\ref{IE(II)}; $N_{\rm A}$ is the Avogadro constant). Close to the surface $\rm S_{12}$ (corresponding to the high values of $\rho$), this surface may be assimilated to its tangent plane, and an approximate solution of the diffusion equation is
\begin{align}
\rho (x,t) &= \frac{j_{\rm m}}{(\pi D)^{1/2}} \int_0^{t-t'} \frac{e^{-x^2/(4Du)}}{u^{1/2}} du \;\;\,{\rm if}\;\,t' < t < t_1,\nonumber\\
\rho (x,t) &= \frac{j_{\rm m}}{(\pi D)^{1/2}} \int_{t-t_1}^{t-t'} \frac{e^{-x^2/(4Du)}}{u^{1/2}} du\;\;\,{\rm if}\;\, t_1 < t \label{rho}
\end{align}
(see Appendix \ref{Diffusion}; $x$ is the distance to $\rm S_{12}$, here assumed small with respect to $r_1$, and $D$ the diffusion coefficient for $\rm O_2$ in water). Note that we have to add to the preceding value the initial constant density $\rho_{\rm a} = c H p_{\rm a}$ of $\rm O_2$ in the solution due to the equilibrium with the atmospheric $\rm O_2$ ($c\, p_{\rm a}$ being the partial pressure of $\rm O_2$ in the atmosphere).

Clearly, at fixed $t$, $\rho$ is a decreasing function of $x$, with a maximum value at $x = 0$
\begin{align}
\rho (0,t) &= \frac{j_{\rm m}}{(\pi D)^{1/2}} \int_0^{t-t'} \frac{du}{u^{1/2}}\nonumber\\ 
&= \frac{2 j_{\rm m}}{(\pi D)^{1/2}} (t-t')^{1/2}\;\;\,{\rm if}\;\,t' < t < t_1,\nonumber\\
\rho (0,t) &= \frac{j_{\rm m}}{(\pi D)^{1/2}} \int_{t-t_1}^{t-t'} \frac{du}{u^{1/2}}\nonumber\\ 
&= \frac{2 j_{\rm m}}{(\pi D)^{1/2}} ((t-t')^{1/2} - (t-t_1)^{1/2})\nonumber\\
&= \frac{2 j_{\rm m}}{(\pi D)^{1/2}}\,\frac{t_1-t'}{(t-t')^{1/2}+(t-t_1)^{1/2}} 
\;\;\,{\rm if}\;\, t_1 < t. 
\end{align}
This explains why bubbles are produced practically at the contact with the nanoelectrode. Moreover, these equations show that $\rho (0,t)$ is an increasing function of $t$ for $t' < t < t_1$ and a decreasing one for $t > t_1$, with a maximum value at $t = t_1$
\begin{eqnarray}
\rho (0,t_1) = \frac{2 j_{\rm m}}{(\pi D)^{1/2}} (t_1-t')^{1/2}. \label{rho max}
\end{eqnarray}

\begin{figure}[htbp]
\begin{center}
\includegraphics[width=6cm]{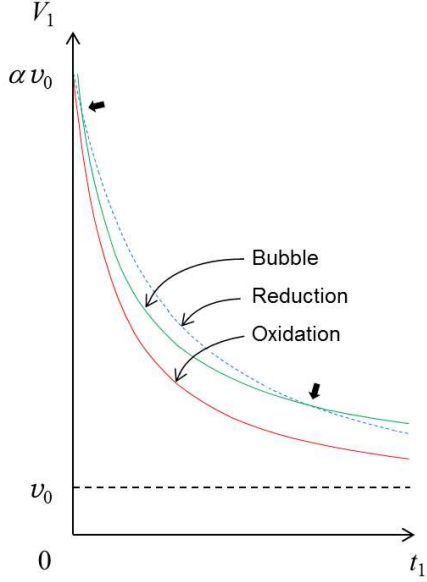}
\end{center}
\caption{In the plane $(t_1,V_1)$, position of the bubble curve Eq.~(\ref{Bubble}) (continuous green curve) with respect to the oxidation curve $t_1 = t'$ (continuous red curve) and the reduction curve $t_1 = \bar t''$ (dashed blue curve) (in zone a or in zone b). The bubble curve intersects the reduction curve at two points (black arrows).} \label{Bubble curve}
\end{figure}

Our simple model is based on this maximum density value: we consider that a bubble will be produced if the maximum value of the density in the solution reaches some supersaturation value $s H p_{\rm a}$
\begin{eqnarray}
\rho (0,t_1) + \rho_{\rm a} = s H p_{\rm a}, \label{s}
\end{eqnarray}
i.e.,
\begin{eqnarray}
(V_1 - v_0)(t_1-t')^{1/2} = 2(s - c)H p_{\rm a}(\pi D)^{1/2} N_{\rm A} q_{\rm e}\frac{\alpha' d}{\gamma}\nonumber\\
\label{Bubble}
\end{eqnarray}
[with the help of Eqs.~(\ref{O2 flux}) and (\ref{rho max})], in which $t'$ is the function of $V_1$ expressed by Eq.~(\ref{t'}) (with $v_0 < V_1 < \alpha\, v_0$). In the plane $(t_1, V_1)$, this ``bubble'' curve [expressed by Eq.~(\ref{Bubble})] is situated above the oxidation curve $t_1 = t'$ but generally intersects (if the coefficient $s$ is not too high) the reduction curve $t_1 = \bar t''$ at two points, so that there is a large part of the bubble curve situated below the reduction curve (see Fig.~\ref{Bubble curve}).
An equation similar to Eq.~(\ref{Bubble}) may be obtained for $\rm H_2$ molecules but the corresponding ``$\rm H_2$ bubble'' curve would be situated above the reduction curve $t_1 = \bar t''$, and then above a large part of the bubble curve Eq.~(\ref{Bubble}): thus, at least in this large part, an $\rm O_2$ bubble will appear but no $\rm H_2$ bubble is produced. In Sec.~\ref{V}, we will see that the whole bubble curve Eq.~(\ref{Bubble}) well represents the experimental bubble curve. In our experiments, with small values of $t_1$ ($0.5\; {\rm ms} < t_1 < 0.5\; {\rm s}$), we generally observe only one bubble, which probably contains both $\rm O_2$ and $\rm H_2$ molecules (in the case $t_1 > \bar t''$). This means that after an $\rm O_2$ bubble is created, the $\rm H_2$ molecules produced during the reduction period $[t_1 + \bar t'',2t_1]$ will enter this bubble (without creating a new bubble).

\begin{figure}[htbp]
\begin{center}
\includegraphics[width=8.6cm]{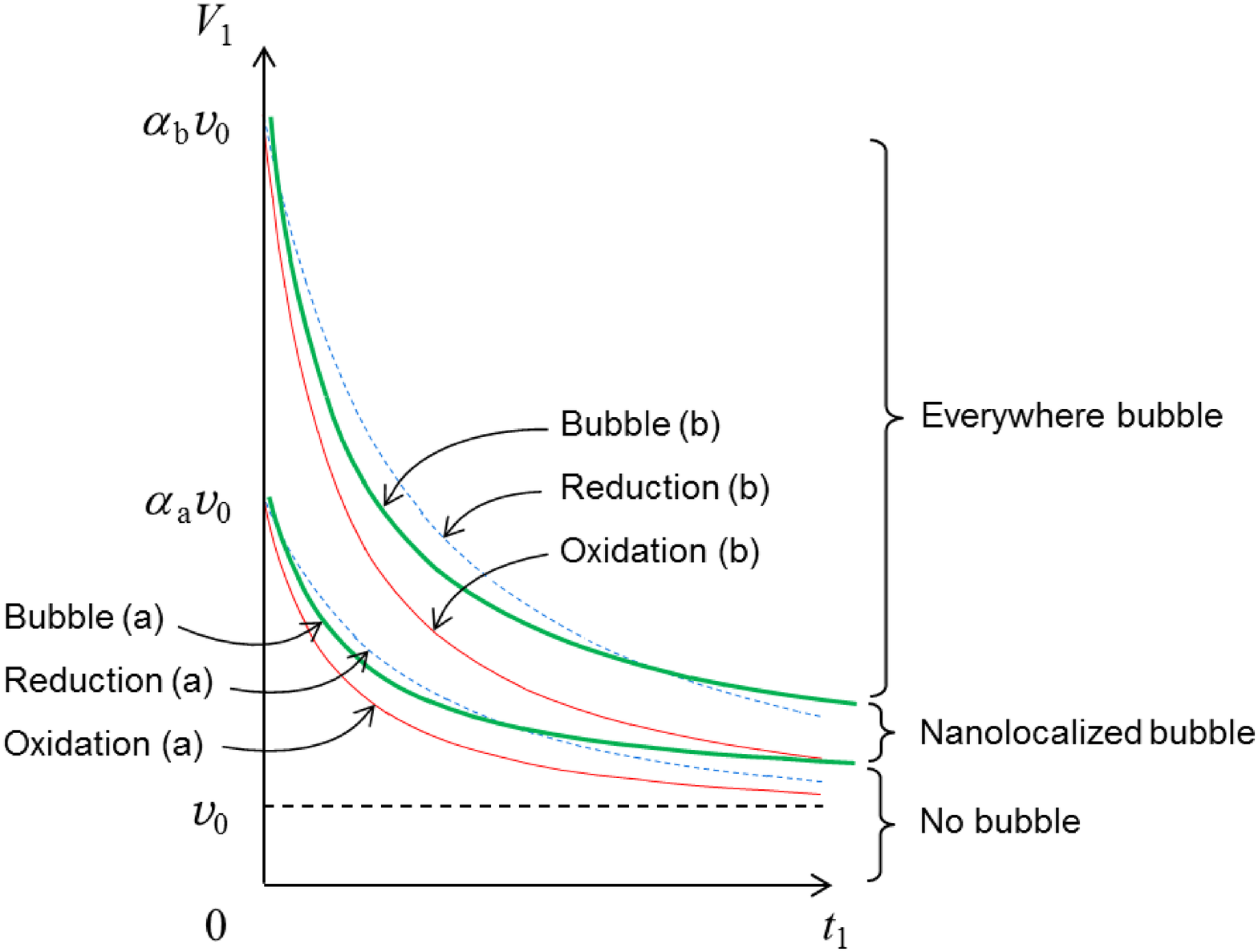}
\end{center}
\caption{In the plane $(t_1,V_1)$, the bubble curve of zone a and that of zone b (thick continuous green curves) delimit the three regions: ``no bubble'' (lower region), ``nanolocalized bubble'' (between the two curves), and ``everywhere bubble'' (upper region).} \label{Bubble ab}
\end{figure}

Clearly, the preceding bubble curve [Eq.~(\ref{Bubble})] delimits two regions: an upper one where bubble production occurs and a lower one with no bubble production. This bubble curve---together with the oxidation and reduction ones---is represented in Fig.~\ref{Bubble ab}, for the zones a and b. Thus, below the bubble curve of zone a, there is no bubble production (neither in zone a nor in zone b). Above the bubble curve of zone b, the bubble production occurs everywhere on the nanoelectrode (in zone a and in zone b). {\it Between the bubble curve of zone a and that of zone b, the bubble production is nanolocalized at the apex of the nanoelectrode} (it occurs in zone a but not in zone b). A numerical example will be given in Sec.~\ref{V}.

\begin{figure}[htbp]
\begin{center}
\includegraphics[width=8.6cm]{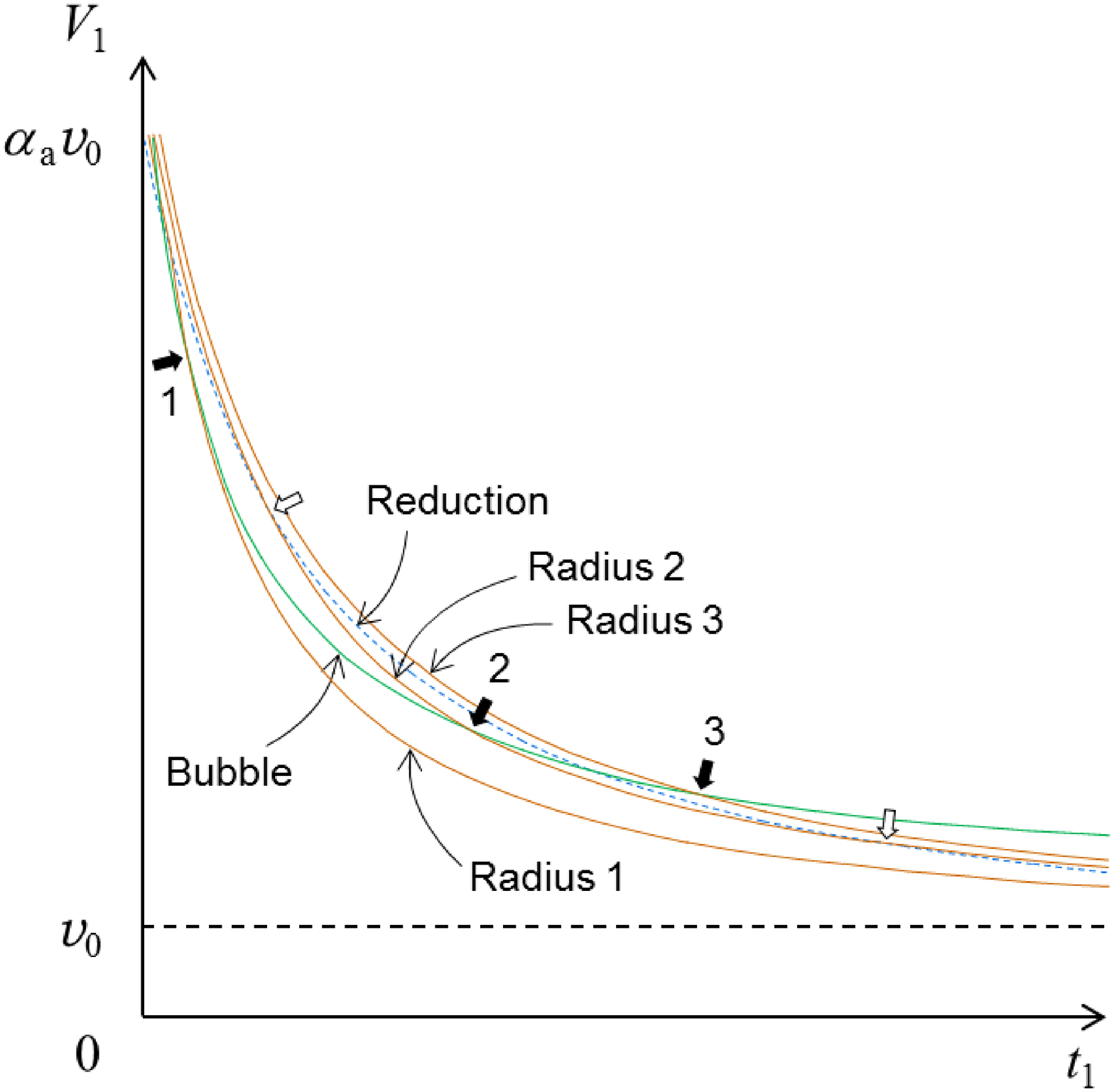}
\end{center}
\caption{In the plane $(t_1,V_1)$, the represented radius 1, radius 2, and radius 3 curves (brown color), respectively, correspond to three increasing radius values, $R_1 < R_2 < R_3$. The two empty arrows indicate the intersections of the radius 2 curve with the reduction curve (dashed blue curve). The black arrows 1, 2, and 3, respectively, indicate the intersections of the bubble curve (green) with the radius 1, radius 2, and radius 3 curves (excepting the region of very small values of $t_1$). As $R$ increases from $R_1$ to $R_3$, we can see that the value of $t_1$ corresponding to these intersection points increases. The oxidation curve (see Fig.~\ref{Bubble curve}), situated below all these curves, is not represented for the sake of clarity.} \label{Radius}
\end{figure}

Let us consider the case of a bubble produced in zone a (i.e., at the apex of the nanoelectrode) and simply assume that when it is formed it contains all the $\rm O_2$ and $\rm H_2$ molecules generated by the electrolysis reactions in zone a. The radius $R$ of the bubble is determined by the number $n$ of these molecules 
\begin{eqnarray}
(p_{\rm a} + \frac{2\sigma}{R})\frac{4}{3}\pi R^3 = n k T \label{n} 
\end{eqnarray}
($p_{\rm a}$ the atmospheric pressure, $\sigma$ the surface tension, $k$ the Boltzmann constant, and $T$ the temperature), which leads to
\begin{align}
(V_1 - v_0)(t_1 - t'_{\rm a}) = \,&\frac{8 q_{\rm e}}{3 k T \gamma r_1}(p_{\rm a} + \frac{2\sigma}{R})R^3 \nonumber\\
&{\rm if}\;\,t'_{\rm a} < t_1 < \bar t''_{\rm a} \label{Radius O2}
\end{align}
(with $n = n_{\rm O_2}$ given by Eqs.~(\ref{nO2}) and (\ref{bar Q'}), with $\alpha'_{\rm a}$ and $a_{\rm a}$; $t'_{\rm a}$ and $\bar t''_{\rm a}$ given by the respective Eqs.~(\ref{t'}) and (\ref{bar t''}), with $\tau_{\rm a}$ and $\alpha_{\rm a}$; $v_0 < V_1 < \alpha_{\rm a}\, v_0$) and to 
\begin{align}
(V_1 - v_0)(3t_1 - t'_{\rm a} - 2\bar t''_{\rm a}) = \,&\frac{8 q_{\rm e}}{3 k T \gamma r_1}(p_{\rm a} + \frac{2\sigma}{R})R^3 \nonumber\\
&{\rm if}\;\,t_1 > \bar t''_{\rm a} \label{Radius O2+H2}
\end{align}
(with $n = n_{\rm O_2} + n_{\rm H_2}$; $n_{\rm O_2}$ as above, $n_{\rm H_2}$ given by Eqs.~(\ref{nH2}) and (\ref{bar Q''}), with $\alpha'_{\rm a}$ and $a_{\rm a}$). In the plane $(t_1,V_1)$, such ``radius'' curves (expressed, for each value of $R$, by Eq.~(\ref{Radius O2}) below the reduction curve and by Eq.~(\ref{Radius O2+H2}) above the reduction curve) are represented in Fig.~\ref{Radius} for three increasing values of $R$, i.e., $R_1$, $R_2$, and $R_3$ with $R_1 < R_2 < R_3$ (respectively, denoted ``radius 1'', ``radius 2'', and ``radius 3'' curves).
They either intersect the reduction curve (of zone a) at two points (indicated by empty arrows for the radius 2 curve in Fig.~\ref{Radius}) or are completely situated above the reduction curve (see the radius 3 curve in Fig.~\ref{Radius}). Moreover, they intersect the above defined bubble curve (of zone a) at one point (if we ignore the very small values of $t_1$ where the bubble curve is above the reduction curve), the corresponding value of $R$ increasing with $t_1$ (see Fig.~\ref{Radius}). Thus, ignoring the very small values of $t_1$ and {\it following the bubble curve of zone a,} i.e., {\it at the limit between the no bubble and the nanolocalized bubble regions, the radius of the bubble (produced at the apex of the nanoelectrode) decreases when $t_1$ decreases}. This corresponds to the intuitive idea that a lower electric charge involved in the electrolysis reactions (then, a lower bubble radius), if released in a shorter time ($t_1 - t'$ decreases if $t_1$ decreases, along the bubble curve), will give the same maximal concentration $\rho(0,t_1)$ in the solution because of diffusion. The above radius curves give the radius of the nanolocalized bubble for each point $(t_1,V_1)$ situated above the bubble curve of zone a (and $V_1 < \alpha_{\rm a}\, v_0$). A numerical example is given in Sec.~\ref{V}.

\section{Discussion} \label{V}

\begin{figure}[htbp]
\begin{center}
\includegraphics[width=5cm]{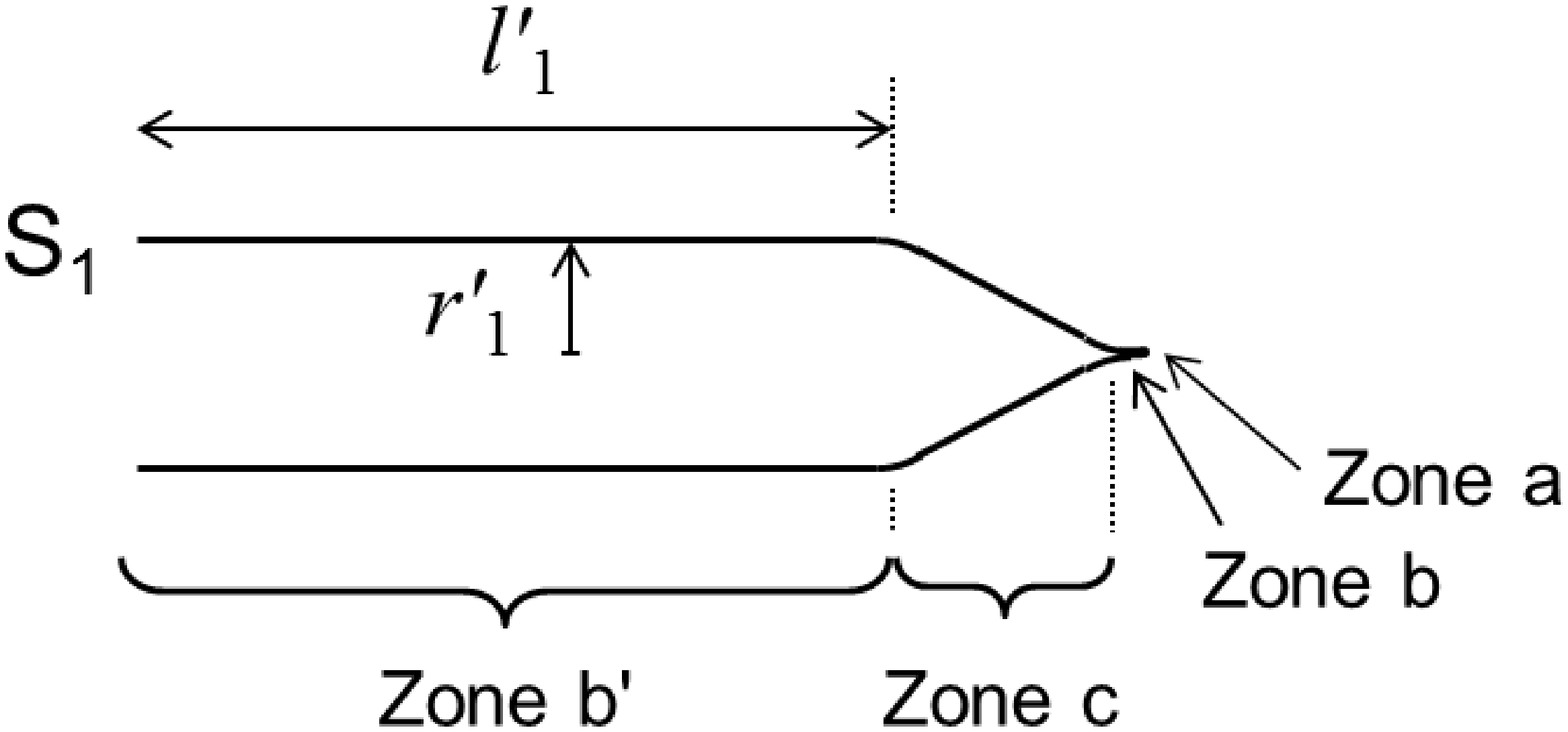}
\end{center}
\caption{Geometry of the whole nanoelectrode surface $\rm S_1$: the zones $\rm a$ and $\rm b$ of Fig.~\ref{S1 geometry}, at the tip, a transition zone $\rm c$, and a long cylinder in zone $\rm b'$.} \label{whole S1}
\end{figure}

In fact, the nanoelectrode geometry shown in Fig.~\ref{S1 geometry} represents the tip of the nanoelectrode. The whole geometry of the nanoelectrode surface $\rm S_1$ may be modelled as (i) the preceding zones $\rm a$ and $\rm b$, at the tip, (ii) a transition zone $\rm c$ (of 100--300 $\mu$m length), and (iii) a long cylinder of length $l'_1$ ($\sim$1 cm) and radius $r'_1$ ($\sim$50 $\mu$m), denoted zone $\rm b'$ (Fig.~\ref{whole S1}).
As above, we assume that the electric potential field in zone $\rm b'$ (except near the two ends of the cylinder) is that produced by a (infinite) cylindrical electrode of radius $r'_1$ and that in zone $\rm c$ is transitional between the values in zones $\rm b$ and $\rm b'$. Clearly, the preceding model for zone $\rm b$ may be applied to zone $\rm b'$, using $r'_1$ and $l'_1$ instead of $r_1$ and $l_1$.

\begin{figure}[htbp]
\begin{center}
\includegraphics[width=6cm]{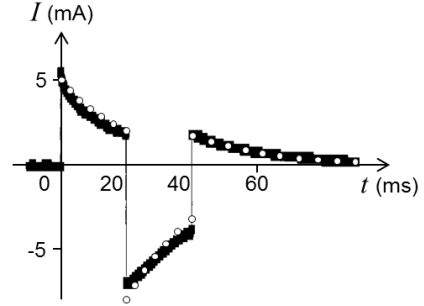}
\end{center}
\caption{Measured current intensity $I$ as a function of time (thick black curve). Potential of rectangular shape with $V_1$ = 35.6 V and $t_1$ = 20 ms, solution with $10^{-3}$ mol/L of $\rm H_2SO_4$ (experiment with nanolocalization of bubble production). The white dots represent the model presented in the paper.} \label{Iexp}
\end{figure}

An experimental curve of the current intensity $I(t)$ is shown in Fig.~\ref{Iexp} (see Refs.~\onlinecite{Hammadi-etal:2013} and \onlinecite{Hammadi-etal:2016} for the experimental procedure; potential of rectangular shape as in Fig.~\ref{V(t)}).
This intensity represents the total intensity originated from the various zones of the nanoelectrode surface. Nevertheless, according to Eq.~(\ref{I}) and owing to the low value of $\beta_2 a$ for zones $\rm a$ and $\rm b$ compared to that of zone $\rm b'$ (e.g., with $r_1$ = 1 $\mu$m, $l_1$ = 10 $\mu$m, $l$ = 1 cm, $r'_1$ = 50 $\mu$m, and $l'_1$ = 1 cm: $\beta_{\rm 2a} a_{\rm a}$ = 6.3 $\mu$m, $\beta_{\rm 2b} a_{\rm b}$ = 6.8 $\mu$m, and $\beta_{\rm 2b'} a_{\rm b'} = 1.2 \times 10^4$ $\mu$m), we may consider that $I$ is mainly originated from zone $\rm b'$ (with some possible contribution of zone $\rm c$). In this experiment, one bubble is produced and nanolocalized at the apex of the nanoelectrode. Assuming no electrolysis reaction in zone $\rm b'$, the above model applied to zone $\rm b'$ gives the current intensity $I(t)$---with the help of Eqs.~(\ref{I}), (\ref{vI}), (\ref{vII}), and (\ref{vIII}), and the discontinuity jumps of $v$ at $t_1$ and $2t_1$---represented by white dots in Fig.~\ref{Iexp}, using the values $\frac{\gamma a}{\alpha' d}(1 - \frac{1}{\alpha})V_1$ = 5 mA and $e^{-t_1/\tau}$ = 0.4. Since $\frac{\gamma a}{\alpha' d}(1 - \frac{1}{\alpha})V_1 \approx \frac{\gamma a}{\alpha d}V_1$ (owing to $\frac{\alpha - 1}{\alpha'} = \beta_1 d \approx 1$), this gives $\frac{a}{\alpha} \approx$ 0.54 $\mu \rm m^2$ (with $d$ = 0.3 nm and $\gamma = 7.8 \times 10^{-2}$ S/m) and $\tau \approx$ 22 ms, i.e., $\alpha = \frac{\gamma}{\varepsilon}\tau \approx 2.4 \times 10^6$ (with $\varepsilon = 80 \,\varepsilon_0$) and $a \approx 1.3 \times 10^6$ $\mu \rm m^2$. These values are of the same magnitude as those corresponding to a cylinder $\rm b'$ of radius $r'_1 \approx$ 50 $\mu$m and length $l'_1 \approx$ 1 cm, i.e., $\alpha \approx 1 + \frac{r'_1}{d}\log \frac{l}{r'_1} \approx 0.9 \times 10^6$ (with $l \approx$ 1 cm) and $a \approx 2\pi r'_1 l'_1 \approx 3 \times 10^6$ $\mu \rm m^2$, which shows that the model is in acceptable agreement with the experimental current intensity data.
 
\begin{figure}[htbp]
\begin{center}
\includegraphics[width=8.6cm]{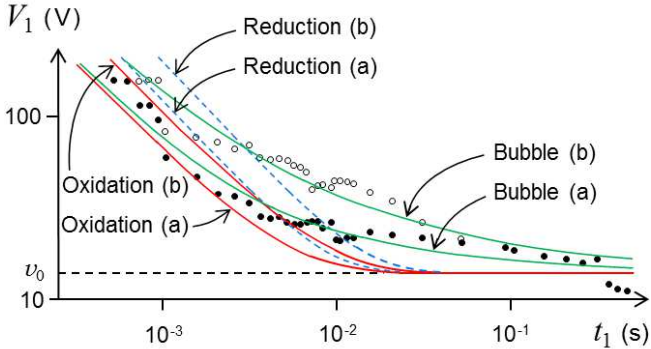}
\end{center}
\caption{Experimental first occurrence of a bubble in zone a (black points) and in zone b (white points), while increasing $V_1$ at constant $t_1$. Solution with $10^{-4}$ mol/L of $\rm H_2SO_4$. These experimental points are, respectively, compared with the bubble curve of zone a [Bubble (a)] and that of zone b [Bubble (b)] of the model (defined by Eq.~(\ref{Bubble}), see Sec.~\ref{IV}; green curves). The oxidation curves (red) and the reduction curves (dashed blue) of the model are also represented (for the zones a and b). Logarithmic scale on both axes.} \label{Bubble exp}
\end{figure}

In a series of experiments at constant $t_1$ and successive increasing values of $V_1$, we note first the experimental value of $V_1$ which corresponds to the first visible occurrence of a bubble in zone a (apex of the electrode, here of micrometric curvature radius) and then the (higher) value of $V_1$ which corresponds to the first visible occurrence of a bubble in zone b (i.e., not at the apex of the electrode). These experimental values are, respectively, plotted as the black and white points in Fig.~\ref{Bubble exp}.
Since the ordinates of the bubble curves (of zone a and of zone b) of the above model tend to $v_0$ when $t_1 \rightarrow +\infty$, we obtain $v_0 \approx$ 14 V from the exprimental points with high values of $t_1$. Such a value of $v_0$ applied to the 0.3 nm thick dielectric film corresponds to a typical ionization field observed in field ion microscopy\cite{Muller-Tsong:1969} and supports the model of electron transfer from water to platinum controlled by field. Moreover, we observed that the radius of the produced unique bubble (in zone a or in zone b, respectively) decreases when $t_1$ decreases (respectively following the black points or the white ones in Fig.~\ref{Bubble exp}). In the experiments with low values of $t_1$, the observed bubble radius is extremely small, which probably means that these points are close to the oxidation curve (of zone a or of zone b; the oxidation curve corresponding to the first occurrence of an electrolysis reaction, when $V_1$ is increased at constant $t_1$). In Fig.~\ref{Bubble exp}, we thus obtain the oxidation curve of zone a (close to the experimental black points for the small values of $t_1$) for a value $r_1 \approx$ 14 $\mu$m [with $v_0$ = 14 V, $d$ = 0.3 nm, $\varepsilon = 80 \,\varepsilon_0$, and $\gamma = 7.8 \times 10^{-3}$ S/m; the oxidation curve being represented by $t_1 = t'$ and $t'$ given by Eq.~(\ref{t'}) with $\alpha_{\rm a}$ and $\tau_{\rm a}$]. It is known that the electric field at the apex of a tip of potential $V$ and curvature radius $r_1$ at its apex (situated at large distance from a counter electrode) is $\approx V/(5\, r_1)$,\cite{Gomer:1961, Gipson-Eaton:1980} i.e., the same as the electric field at the surface of a sphere of potential $V$ and radius $\tilde r_1 \approx 5\, r_1$. The potential $v$ being directly related to this electric field [$E_{1\rm n} \approx v/d$, Eqs.~(\ref{E1n-E2n}) and (\ref{beta1a-beta2a})], our simple model of a sphere for zone a should be applied with a sphere radius $\tilde r_1 \approx 5\, r_1$ ($r_1$ being the curvature radius at the apex of the electrode). The above obtained value $\tilde r_1 \approx$ 14 $\mu$m thus corresponds to a curvature radius $r_1 \approx \tilde r_1/5 \approx$ 2.8 $\mu$m at the apex, in qualitative agreement with the micrometric curvature radius at the apex of our electrode. Moreover, this value $r_1 =$ 2.8 $\mu$m (with $l =$ 1 cm) leads to the oxidation curve of zone b represented in Fig.~\ref{Bubble exp} [i.e., $t_1 = t'$ and $t'$ given by Eq.~(\ref{t'}) with $\alpha_{\rm b}$ and $\tau_{\rm b}$] which, as expected and noted earlier, is close to the experimental white points for the small values of $t_1$. The reduction curves of zones a and b [i.e., $t_1 = \bar t''$ and $\bar t''$ given by Eq.~(\ref{bar t''}), using $\tilde r_1$ for zone a and $r_1$ for zone b] are also represented in Fig.~\ref{Bubble exp}. Note that, because of this value $\tilde r_1$ used in zone a (higher than the value $r_1$ of zone b), the reduction curve of zone a is close to the oxidation curve of zone b, for the values of $t_1$ represented in Fig.~\ref{Bubble exp}.

The bubble curves of zones a and b represented in Fig.~\ref{Bubble exp} [i.e., Eq.~(\ref{Bubble}), using $\tilde r_1$ for zone a and $r_1$ for zone b; $c$ = 0.21, $H$ = 1.3 mol/($\rm m^3$ atm) and $D = 2.5 \times 10^{-9}$ $\rm m^2$/s at $25^{\circ}$C], with a coefficient $s$ = 20 for zone a and $s$ = 30 for zone b, show a rather good agreement with the exprimental points (i.e., the black points for the bubble curve of zone a and the white points for the bubble curve of zone b). Note that the production of a visible bubble is based on its nucleation and growth up to a micrometric size, which requires that a sufficient domain (in space and time) of the solution becomes supersaturated. This probably explains the obtained high values $s$ = 20--30 which, according to Eq.~(\ref{s}), correspond to the maximum supersaturation of the solution at only the single point $(x,t) = (0,t_1)$.

\begin{figure}[htbp]
\begin{center}
\includegraphics[width=8.6cm]{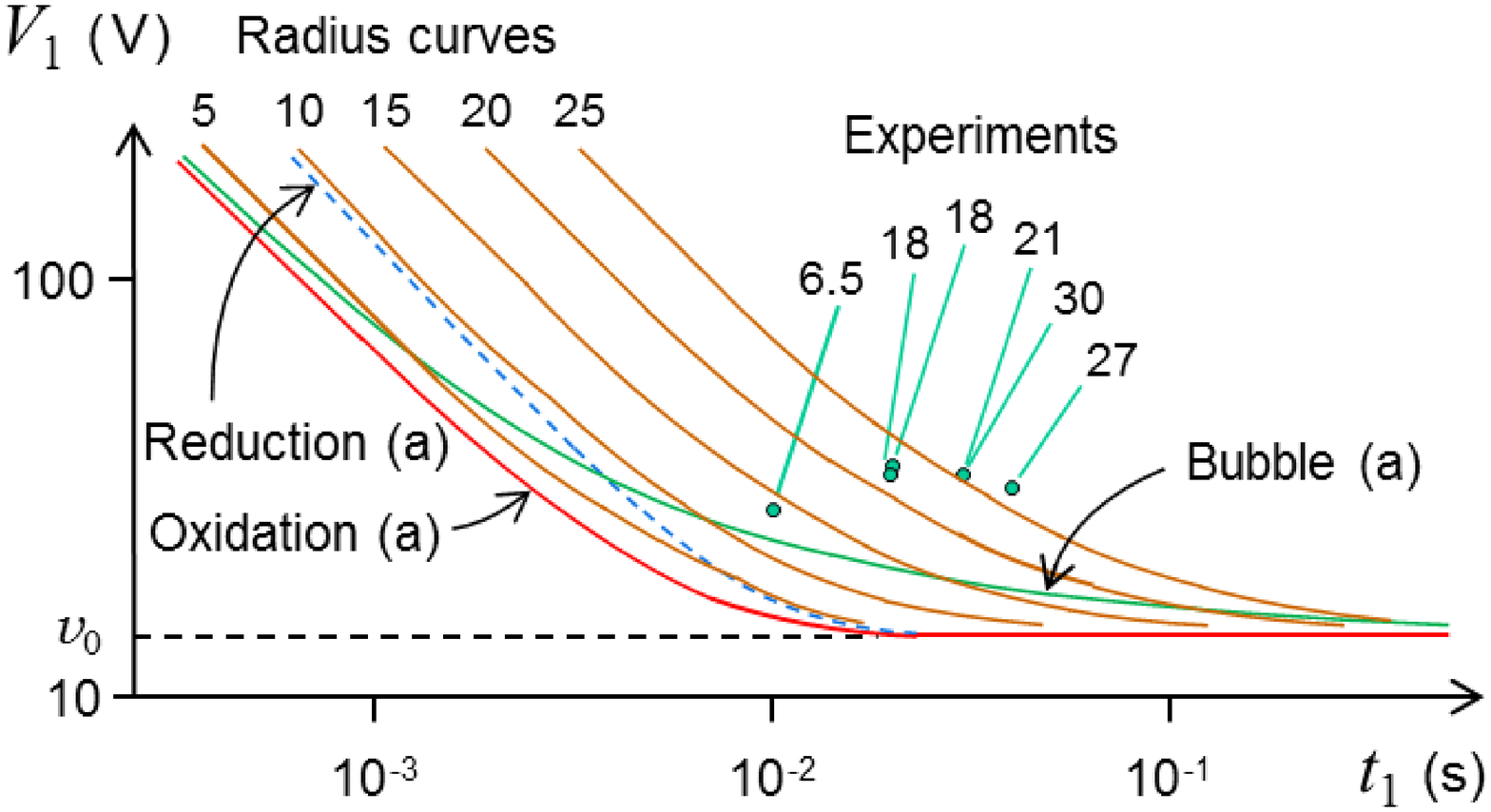}
\end{center}
\caption{Measured radii of nanolocalized bubbles (experimental green points; for each experiment, the number indicates the measured radius in $\mu$m). Solution with $10^{-4}$ mol/L of $\rm H_2SO_4$. These values may be compared with the radius curves of the model presented in the paper (brown curves; above each curve, the number indicates the radius value in $\mu$m). The oxidation (red), reduction (dashed blue), and bubble (green) curves of the model are also represented (for zone a). Logarithmic scale on both axes.} \label{Radius exp}
\end{figure}

In some experiments with similar conditions (same type of electrode, solution with $10^{-4}$ mol/L of $\rm H_2SO_4$), the initial radius of the nanolocalized bubble (i.e., the unique bubble produced at the apex of the electrode; this radius then tends to decrease because the solution becomes undersaturated and the bubble dissolves in the solution) was measured (green points in Fig.~\ref{Radius exp}). These measurements are in qualitative agreement with the radius curves of the present model represented in Fig.~\ref{Radius exp} [i.e., for each radius value $R$, Eq.~(\ref{Radius O2}) below the reduction curve of zone a and Eq.~(\ref{Radius O2+H2}) above this reduction curve, using the above values of $v_0$, $d$, $\varepsilon$, $\gamma$, $\tilde r_1$, and $\sigma = 7.2 \times 10^{-2}$ N/m at $25^{\circ}$C]. This confirms that the nanolocalized bubble is approximately only formed by the ${\rm O_2}$ and ${\rm H_2}$ molecules produced in zone a (and not those produced in zone b, the experimental points of Fig.~\ref{Radius exp} being situated above the oxidation and reduction curves of zone b shown in Fig.~\ref{Bubble exp}). Also in agreement with the observations, and as mentioned at the end of Sec.~\ref{IV}, the model shows that, following the bubble curve of zone a, in Fig.~\ref{Radius exp}, the bubble radius decreases when $t_1$ decreases.

\section{Conclusion}

We define nanoelectrolysis as the nanolocalization (at a single point) of electrolysis phenomena such as the electrolysis reactions or the production of bubbles. This paper presents a simple model of water nanoelectrolysis based on (i) the higher value of the electric field at the apex of a tip-shaped electrode (called the nanoelectrode; 5 $\mu$m to 1 nm of curvature radius at the apex) and (ii) the electron tunneling assisted by the electric field through the thin film of water molecules ($\sim$0.3 nm thick) at the surface of the electrode. Using a finite time $t_1$ (related to the amplitude of the potential and to the geometry of the nanoelectrode) during which a positive (respectively, a negative) electric potential $V_1$ is applied between the electrodes, we show that the electrolysis oxidation reaction which produces ${\rm O_2}$ molecules (respectively, the reduction reaction which produces ${\rm H_2}$ molecules) can be nanolocalized at the apex of the nanoelectrode. These two parameters $t_1$ and $V_1$ determine the time at which the electrolysis reaction begins, the electrolysis current intensity (i.e., the tunneling current intensity), the electric charge delivered by this current, and then the number of produced ${\rm O_2}$ or ${\rm H_2}$ molecules (${\rm O_2}$ if $V_1 > 0$, ${\rm H_2}$ if $V_1 < 0$). It is also shown that, during the electrolysis reaction (oxidation or reduction), the current intensity is equal to the electrolysis current intensity (assuming a high ``tunneling conductivity'' $\gamma_0 = dj/dE$ above a threshold value $E_0$ of the electric field). The model then determines the concentration of ${\rm O_2}$ or ${\rm H_2}$ molecules in the solution close to the surface of the nanoelectrode (as a consequence of the electrolysis current and the diffusion of these molecules in the solution). A bubble is then produced if the maximum of this concentration reaches some supersaturation value.

This general model is applied to the simple case of a potential (between the electrodes) of rectangular shape: $V(t) = V_1$ for $0 < t < t_1$, $V(t) = -V_1$ for $t_1 < t < 2t_1$, and $V(t) = 0$ for $t < 0$ and $t > 2t_1$. If, e.g., $V_1 > 0$, it is shown that the duration of the electrolysis reduction reaction is lower than that of the oxidation reaction (when these two reactions occur) and the number of produced $\rm H_2$ molecules is lower than twice the number of produced $\rm O_2$ molecules (if a unique threshold value $E_0$ of the electric field is used\cite{Asymmetry}). Moreover, in the plane $(t_1,V_1)$, the region for the nanolocalization of the reduction reaction is shifted toward higher values of $t_1$ or $V_1$, with respect to that for the nanolocalization of the oxidation reaction. In the same plane, we show that there is a third distinct region, that for the nanolocalization of the production of bubbles (also shifted toward higher values of $t_1$ or $V_1$, with respect to that for the nanolocalization of the oxidation reaction). From the number of produced ${\rm O_2}$ and ${\rm H_2}$ molecules, the model also determines the radius of the produced (unique) nanolocalized bubble, as a function of $(t_1,V_1)$. In the plane $(t_1,V_1)$, following the ``bubble curve of the apex zone'', i.e., at the limit between the no bubble and the nanolocalized bubble regions, the radius of the nanolocalized bubble decreases when $t_1$ decreases.

In addition to the nanolocalization of the production of bubbles (first observed in Ref.~\onlinecite{Hammadi-etal:2013}), the model also explains why there may be nanolocalization of an electrolysis reaction (oxidation or reduction) whereas no bubble is produced (as was assumed in our previous paper about the immobilization of a bubble\cite{Hammadi-etal:2016}). Moreover, the above results of the model are in rather good agreement with our recent experiments with a potential of rectangular shape, concerning the current intensity, the region for the nanolocalization of the production of bubbles [in the plane $(t_1,V_1)$], and the radii of the nanolocalized bubbles. We also currently try to measure the current intensity originated only from the apex of the nanoelectrode, which, according to the present model, is equal to the electrolysis current intensity (i.e., the tunneling current; during the electrolysis reaction). In conclusion, the model shows that, by means of the two parameters $t_1$ and $V_1$, we can control (i) the nanolocalization of the oxidation reaction, (ii) the nanolocalization of the reduction reaction, or (iii) the nanolocalization of the production of bubbles, but also the time at which the electrolysis reaction (of oxidation or reduction) begins, the duration of this reaction, the electrolysis current intensity, the number of produced ${\rm O_2}$ or ${\rm H_2}$ molecules, and the radius of the nanolocalized bubble. 

\appendix

\section{Spherical and cylindrical electrodes} \label{Sph Cyl}

\subsection{Spherical electrodes}

In this case, $\rm S_1$, $\rm S_{12}$, and $\rm S_2$ are concentric spheres, of respective radii $r_1$, $r_1 + d$, and $r_1 + d + l$ (in fact, if $l \gg r_1 + d$, the form of $\rm S_2$ has no significant effect on the potential field in the neighbourhood of $\rm S_1$ and $\rm S_{12}$). We may consider that the potential is zero on $\rm S_2$. According to Maxwell's equations, at a given time $t$, the potential $\varphi$ satisfies
\begin{eqnarray*}
\Delta \varphi = \frac{d^2 \varphi}{d r^2} + \frac{2}{r}\frac{d \varphi}{d r}
= \frac{1}{r}\frac{d^2 (r\varphi)}{d r^2} = 0
\end{eqnarray*}
($r$ being the distance to the centre of the spheres), hence
\begin{eqnarray*}
\frac{d (r\varphi)}{d r} = \varphi + r \frac{d \varphi}{d r} = a
\end{eqnarray*}
($a$ constant; thus, $r\varphi = a r + b$, with $b$ constant, i.e., $\varphi = a + \frac{b}{r}$). The preceding equation applied in region $\rm V_1$ between $\rm S_1$ and $\rm S_{12}$
\begin{eqnarray*}
(\varphi + r \frac{d \varphi}{d r})_1 = \frac{\delta (r\varphi)}{\delta r}
\end{eqnarray*}
(the subscript 1 indicates the value on the $\rm V_1$ side of $\rm S_{12}$, and $\delta$ the value on $\rm S_{12}$ minus that on $\rm S_1$), i.e.,
\begin{eqnarray*}
\varphi_{12} + (r_1 + d) (\frac{d \varphi}{d r})_1 
= \frac{(r_1 + d)\varphi_{12} - r_1 \varphi_1}{d}
\end{eqnarray*}
($\varphi_1 = V$ and $\varphi_{12} = V - v$, respectively, being the potentials on $\rm S_1$ and $\rm S_{12}$), leads to
\begin{eqnarray*}
E_{1\rm n} = -(\frac{d \varphi}{d r})_1 
= \frac{r_1(\varphi_1 - \varphi_{12})}{(r_1 + d)d} = \frac{r_1}{(r_1 + d)d}\; v.
\end{eqnarray*}

Similarly, the same equation applied in region $\rm V_2$ between $\rm S_{12}$ and $\rm S_2$
\begin{eqnarray*}
(\varphi + r \frac{d \varphi}{d r})_2 = \frac{\delta (r\varphi)}{\delta r}
\end{eqnarray*}
(the subscript 2 indicates the value on the $\rm V_2$ side of $\rm S_{12}$, and $\delta$ the value on $\rm S_2$ minus that on $\rm S_{12}$), i.e.,
\begin{eqnarray*}
\varphi_{12} + (r_1 + d) (\frac{d \varphi}{d r})_2 
= \frac{-(r_1 + d)\varphi_{12}}{l}
\end{eqnarray*}
leads to
\begin{eqnarray*}
E_{2\rm n} = -(\frac{d \varphi}{d r})_2 
= \frac{(r_1 + d + l)\varphi_{12}}{(r_1 + d)l} = \frac{r_1 + d + l}{(r_1 + d)l}\; (V - v).
\end{eqnarray*}

\subsection{Cylindrical electrodes}

Here, $\rm S_1$, $\rm S_{12}$, and $\rm S_2$ are coaxial (infinite) cylinders, of respective radii $r_1$, $r_1 + d$, and $r_1 + d + l$ (here also, if $l \gg r_1 + d$, the form of $\rm S_2$ has no significant effect on the potential field in the neighbourhood of $\rm S_1$ and $\rm S_{12}$). We may consider that the potential is zero on $\rm S_2$. According to Maxwell's equations, at a given time $t$, the potential $\varphi$ satisfies
\begin{eqnarray*}
\Delta \varphi = \frac{d^2 \varphi}{d r^2} + \frac{1}{r}\frac{d \varphi}{d r}
= \frac{1}{r}\frac{d}{d r}(r \frac{d \varphi}{d r}) = 0
\end{eqnarray*}
($r$ being the distance to the axis of the cylinders), hence
\begin{eqnarray*}
r \frac{d \varphi}{d r} = \frac{d \varphi}{d \log r} = a
\end{eqnarray*}
($a$ constant; thus, $\varphi = a \log r + b$, with $b$ constant). The preceding equation applied in region $\rm V_1$ between $\rm S_1$ and $\rm S_{12}$
\begin{eqnarray*}
(r \frac{d \varphi}{d r})_1 = \frac{\delta \varphi}{\delta \log r}
\end{eqnarray*}
(same notations as above, for spherical electrodes), i.e.,
\begin{eqnarray*}
(r_1 + d) (\frac{d \varphi}{d r})_1 
= \frac{\varphi_{12} - \varphi_1}{\log\frac{r_1 + d}{r_1}}
\end{eqnarray*}
leads to
\begin{eqnarray*}
E_{1\rm n} = -(\frac{d \varphi}{d r})_1 
= \frac{v}{(r_1 + d)\log\frac{r_1 + d}{r_1}}.
\end{eqnarray*}

Similarly, the same equation applied in region $\rm V_2$, between $\rm S_{12}$ and $\rm S_2$,
\begin{eqnarray*}
(r \frac{d \varphi}{d r})_2 = \frac{\delta \varphi}{\delta \log r}
\end{eqnarray*}
(same notations as for spherical electrodes), i.e.,
\begin{eqnarray*}
(r_1 + d) (\frac{d \varphi}{d r})_2 
= \frac{-\varphi_{12}}{\log\frac{r_1 + d + l}{r_1 + d}},
\end{eqnarray*}
leads to
\begin{eqnarray*}
E_{2\rm n} = -(\frac{d \varphi}{d r})_2 
= \frac{V - v}{(r_1 + d)\log\frac{r_1 + d + l}{r_1 + d}}.
\end{eqnarray*}

\section{Proof of $v(2t_1^+) < 0$} \label{vIII<0}

\subsection{Case $t_1 < t'$}

According to Eqs.~(\ref{vI}) and (\ref{vII}), and the discontinuity jumps at $t_1$ and $2t_1$, we obtain
\begin{eqnarray*}
v(2t_1^+) = -(V_1 - \frac{V_1}{\alpha})(1 - e^{-t_1/\tau})^2 < 0.
\end{eqnarray*}

\subsection{Case $t' < t_1 < t''$}

According to Eqs.~(\ref{vIE}) and (\ref{vII}), and the discontinuity jumps at $t_1$ and $2t_1$, we obtain
\begin{align*}
v(2&t_1^+) = -V_1 + \frac{V_1}{\alpha}\\ 
&+ (V_1 + V_{1+} - 2\frac{V_1}{\alpha} - (V_{1+} - v_0)e^{-(t_1 - t')/\tau'})e^{-t_1/\tau},
\end{align*}
which is a function of $t_1$ (at $V_1$ constant). Then
\begin{align*}
&\frac{d(v(2t_1^+))}{dt_1} = \frac{e^{-t_1/\tau}}{\tau} \times\\ 
&((V_1 - v_0 + V_{1+} - v_0)e^{-(t_1 - t')/\tau'} - V_1 - V_{1+} + 2\frac{V_1}{\alpha})\\
&< -\frac{e^{-t_1/\tau}}{\tau} 2(v_0 - \frac{V_1}{\alpha}) < 0
\end{align*}
(using $e^{-(t_1 - t')/\tau'} < 1$), which shows that $v(2t_1^+)$ is a decreasing function of $t_1$. Thus, $v(2t_1^+)$ (at $t_1$) is lower than its value at $t_1 = t'$ (since $t_1 > t'$)
\begin{align*}
&v(2t_1^+) < -(1 - \frac{1}{\alpha})V_1 + ((1 - \frac{2}{\alpha})V_1 + v_0)e^{-t'/\tau}\\
&= -\frac{(v_0 - \frac{V_1}{\alpha})^2}{(1 - \frac{1}{\alpha})V_1} < 0
\end{align*}
[using the value of $t'$ given by Eq.~(\ref{t'})].

\subsection{Case $t'' < t_1$}

In this case, $v(2t_1^-) < -v_0$, then $v(2t_1^+) < -v_0 + \frac{V_1}{\alpha} < 0$.

\section{Solution of the diffusion equation} \label{Diffusion}

The one-dimensional diffusion equation
\begin{eqnarray*}
\frac{\partial \rho}{\partial t}(x,t) - D \frac{\partial^2 \rho}{\partial x^2}(x,t) = \delta_0(x) \delta_{t_0}(t)
\end{eqnarray*}
($\delta_u$ denotes the Dirac measure at $u$) has the known elementary solution in ${\bf R} \times ]t_0,+\infty[$
\begin{eqnarray*}
\rho (x,t) = \frac{e^{-x^2/(4D(t-t_0))}}{(4\pi D (t-t_0))^{1/2}}.
\end{eqnarray*}
In our case, the surface $\rm S_{12}$ being assimilated to the plane $x = 0$ and the aqueous solution occupying the region $x > 0$, $\rho$ does not depend on the other coordinates $y$ and $z$, and the elementary problem of diffusion
\begin{eqnarray*}
\frac{\partial \rho}{\partial t}(x,t) - D \frac{\partial^2 \rho}{\partial x^2}(x,t) = j_{\rm m}\, dt_0\, \delta_0(x) \delta_{t_0}(t)
\end{eqnarray*}
has the similar solution in ${\bf R_{\scriptscriptstyle +}^*} \times ]t_0,+\infty[$
\begin{eqnarray*}
\rho (x,t) = 2 j_{\rm m}\, dt_0 \frac{e^{-x^2/(4D(t-t_0))}}{(4\pi D (t-t_0))^{1/2}},
\end{eqnarray*}
with the coefficient 2 because the produced molecules diffuse in the half space $x > 0$. The integration on $t_0 \in [t',t_1]$ then leads to Eq.~(\ref{rho}) for $t_1 \leq t$. The case $t' < t < t_1$ is simply obtained by applying this equation at $t = t_1$ and then replacing $t_1$ with $t$.

\bibliography{Nanoelectrolysis}

\begin{thebibliography}{33}
\bibitem [{Asy()}]{Asymmetry}%
  \BibitemOpen
  \href@noop {} {}\bibinfo {note} {The solution-to-metal electron tunneling
  being not strictly equivalent to the metal-to-solution one, distinct
  threshold values (and slopes) may be used: $E_0^+ = v_0^+/d$ and $\gamma_0^+$
  for $E > 0$, and $E_0^- = v_0^-/d$ and $\gamma_0^-$ for $E < 0$. It is then
  easy to replace, in the following, $v_0$ and $\gamma_0$ with $v_0^+$ and
  $\gamma_0^+$, if $v > 0$, and with $v_0^-$ and $\gamma_0^-$, if $v <
  0$.}\BibitemShut {Stop}%
\end{thebibliography}
 
\end{document}